\documentclass[10pt,twocolumn,letterpaper]{article}

\usepackage{iccv}
\usepackage{times}
\usepackage{epsfig}
\usepackage{graphicx}
\usepackage{amsmath}
\usepackage{amssymb}

\usepackage{multirow}
\usepackage[american]{babel}
\usepackage{array}
\usepackage{booktabs}
\usepackage[abs]{overpic}
\usepackage{nccmath}
\usepackage{bbding}
\usepackage[accsupp]{axessibility}

\usepackage[pagebackref=true,breaklinks=true,letterpaper=true,colorlinks,bookmarks=false]{hyperref}

\usepackage[capitalize]{cleveref}
\crefname{section}{Sec.}{Secs.}
\Crefname{section}{Section}{Sections}
\Crefname{table}{Table}{Tables}
\crefname{table}{Tab.}{Tabs.}
\Crefname{equation}{Equation}{Equations}
\crefname{equation}{Eqn.}{Eqns.}

\newcommand{\lxm}[1]{\textcolor[rgb]{0,0,0}{#1}}

\iccvfinalcopy 

\def\iccvPaperID{****} 

\ificcvfinal\fi

\begin{document}

\title{Beyond Static Scenes: Camera-controllable Background Generation \\for Human Motion}

\author{Mingshuai Yao$^{1,2}$, Mengting Chen$^{2(}$\dag$^)$, Qinye Zhou$^2$, Yabo Zhang$^1$, Ming Liu$^1$, Xiaoming Li$^{1(}$\Envelope$^)$, \\Shaohui Liu$^1$ , Chen Ju$^2$, Shuai Xiao$^2$, Qingwen Liu$^2$, Jinsong Lan$^{2(}$\Envelope$^)$, Wangmeng Zuo$^1$\\
$^1$Harbin Institute of Technology, Harbin, China \quad $^2$ Taobao and Tmall Group\\
{\tt\small 
\href{mailto:ymsoyosmy@gmail.com}{\color{black}ymsoyosmy@gmail.com}, \href{mailto:csmliu@outlook.com}{\color{black}csmliu@outlook.com}, \href{mailto:wmzuo@hit.edu.cn}{\color{black}wmzuo@hit.edu.cn}}
}

\maketitle
\ificcvfinal\thispagestyle{empty}\fi

\begin{figure*}[t]
	\setlength{\abovecaptionskip}{7pt} 
	\setlength{\belowcaptionskip}{-10pt}
	\centering
	\includegraphics[width=.99\textwidth]{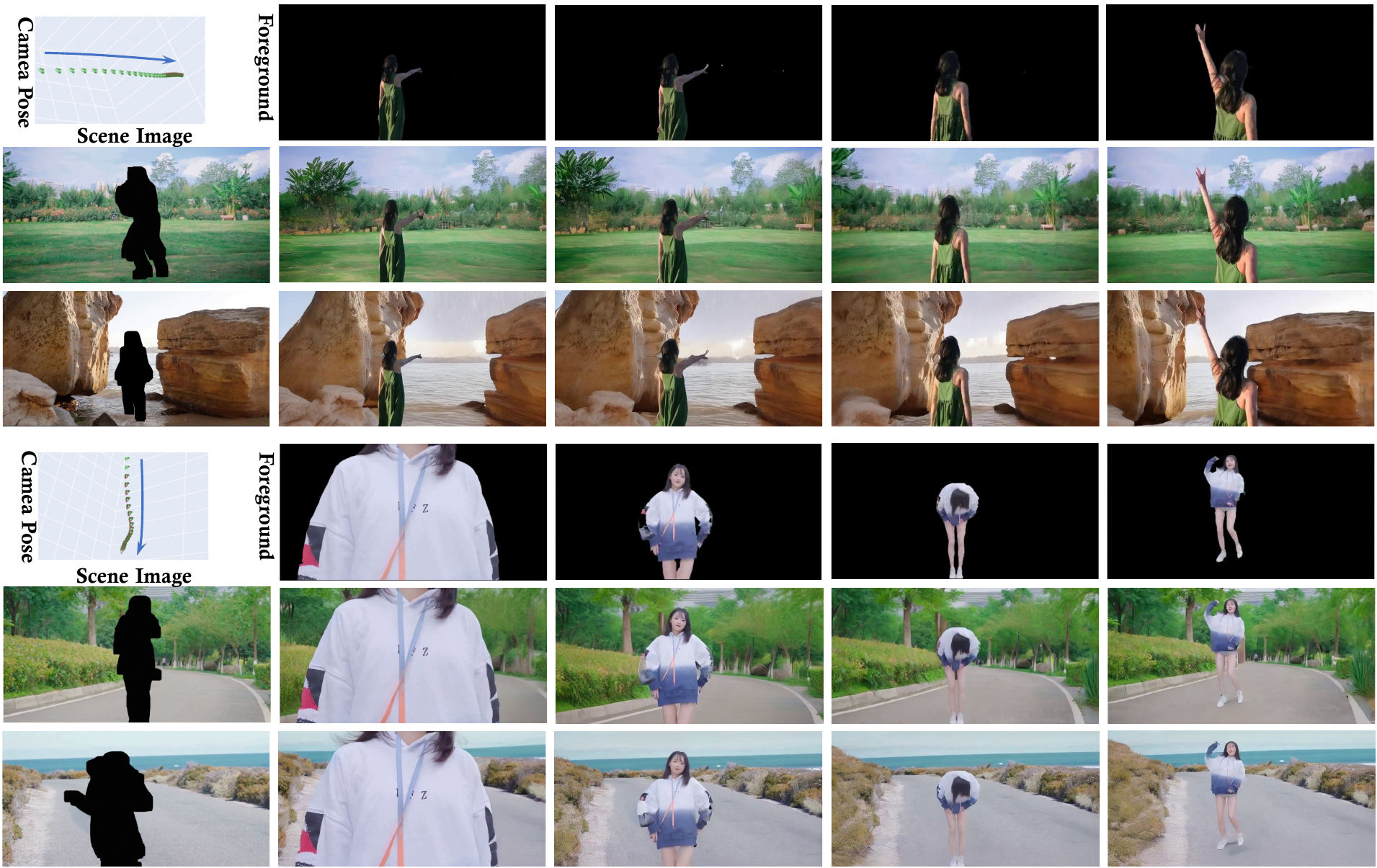}
	\caption{
    Results of DynaScene on different camera motion trajectories and backgrounds.
    The first and fourth rows are the video foreground, with the scene image in the first column, followed by the four generated results. Our approach not only allows the video foreground to move seamlessly in any given scene but also ensures that the background adapts to the motion of the foreground human.
 }
	\label{fig:pipeline_b} 
\end{figure*}
\begin{abstract}
\noindent In this paper, we investigate the generation of new video backgrounds given a human foreground video, a camera pose, and a reference scene image. This task presents three key challenges. First, the generated background should precisely follow the camera movements corresponding to the human foreground. Second, as the camera shifts in different directions, newly revealed content should appear seamless and natural. Third, objects within the video frame should maintain consistent textures as the camera moves to ensure visual coherence. To address these challenges, we propose \textbf{DynaScene}, a new framework that uses camera poses extracted from the original video as an explicit control to drive background motion. Specifically, we design a multi-task learning paradigm that incorporates auxiliary tasks, namely background outpainting and scene variation, to enhance the realism of the generated backgrounds. Given the scarcity of suitable data, we constructed a large-scale, high-quality dataset tailored for this task, comprising video foregrounds, reference scene images, and corresponding camera poses. 
This dataset contains 200K video clips, \lxm{ten times larger than existing real-world human video datasets}, providing a significantly richer and more diverse training resource. Project page: \href{https://yaomingshuai.github.io/Beyond-Static-Scenes.github.io/}{\color{black}https://yaomingshuai.github.io/Beyond-Static-Scenes.github.io/}
\end{abstract}

\vspace{-20pt}
\footnote{$\dag$ Project Lead}
\section{Introduction}
\label{sec:intro}
Camera-controllable background generation enables dynamic scene compositions that maintain spatiotemporal coherence with both subject movements and viewpoint changes, allowing for more realistic scene composition.
However, manual creation of such synchronized background presents a laborious and time-consuming bottleneck in content creation workflows.
Currently, most video editing models focus on foreground motion manipulation~\cite{wang2023disco, chang2023magicdance,feng2023dreamoving,xu2024magicanimate,hu2024animate,zhu2024champ,tan2024animate,wang2025cinemaster} or global camera motion synthesis~\cite{wang2024motionctrl,he2024cameractrl,wang2024humanvid},  while the critical challenge of camera-aware background generation remains largely unexplored.

A notable attempt to address this issue is ActAnywhere~\cite{pan2024actanywhere}, which pioneers video background generation using diffusion models~\cite{ho2020denoising,song2020denoising}. 
By conditioning background synthesis on foreground videos and static scene images, ActAnywhere generates scene-aware backgrounds.
However, it struggles to explicitly model camera motion dynamics, 
as the background generation relies solely on foreground-driven cues.
Consider a scenario where a person walks forward at a velocity matching the camera’s dolly-in motion, 
or both the person and the camera move to the right,
creating the illusion that the subject remains relatively still in the frame.
Since ActAnywhere~\cite{pan2024actanywhere} relies only on foreground motion for background generation, it fails to account for camera movement, leading to inconsistent or unnatural background motion. 
\lxm{This highlights the necessity of providing explicit camera poses to ensure accurate and consistent background motion.}
Meanwhile, recent text-to-video synthesis methods~\cite{wang2024motionctrl,he2024cameractrl,wang2024humanvid}  introduce parametric control over camera trajectories through explicit pose conditioning.
Building on these insights, we propose a new framework that explicitly couples camera motion control with background generation via dedicated kinematic constraints. Given a human foreground sequence, a reference scene image, and camera pose parameters, we synthesize dynamic backgrounds that naturally adapt to the foreground subject’s movement while ensuring precise consistency with camera motion.
This physically grounded interaction between foreground and background enables the creation of immersive video content with synchronized spatiotemporal evolution.

However, existing datasets fall short for training effective camera-controllable video background generation models.
The Realestate10K dataset~\cite{zhou2018stereo}, 
primarily consists of scenery videos with camera movement but lacks any human foregrounds.
This limitation restricts its applicability to tasks involving foreground motion and background interaction. Additionally, many popular open-source human datasets~\cite{zhou2018stereo, zablotskaia2019dwnet, black2023bedlam, lin2024motion, wang2024humanvid} either have static backgrounds or exhibit only minimal camera movements. Consequently, these datasets are not well-suited for training models that require more dynamic background generation.
To bridge this gap, we introduce a new large-scale dataset specifically designed for camera-controllable video background generation. It contains over 200,000 video clips, each with diverse and dynamic camera movement. These clips are paired with corresponding scene images and camera poses, all in high-quality 1080P resolution. The dataset includes a wide variety of dynamic human activities, such as dancing, sports, photography teaching, skateboarding, skiing, and parkour, offering a rich variety of human foregrounds and complex background motions. This large-scale and diverse dataset provides a solid foundation for training models capable of generating video backgrounds that can dynamically adapt to both human movement and camera poses.

As the camera moves, both newly revealed areas and objects already present in the frame should maintain consistent textures and structural coherence in the generated background. 
To achieve this, we integrate multi-task learning into our approach, using background outpainting and scene variation techniques. This enables the model to not only synthesize contextual objects with realistic details but also adapt to different perspectives as the camera moves.
With this dataset and methodology, our approach ensures that the foreground can move freely within any given scene while the background evolves dynamically in response.
As shown in Figure~\ref{fig:pipeline_b}, as the foreground characters move, the generated background remains spatial consistency, preserving both global scene structure and local texture details.
The contributions of our work are summarized as follows:
\begin{itemize}
\item We propose a new framework for video background generation that is explicitly controlled by camera poses. To support this, we introduce a large-scale real-world dataset of 200K video clips \lxm{(10$\times$ larger than others)} featuring dynamic human action and diverse camera movements.
\item We present a multi-task learning strategy that integrates background outpainting and scene variation, enabling the generation of content-consistent backgrounds that seamlessly adapt to camera motion and foreground movement.
\end{itemize}
\section{Related Work}
\label{sec:Controllable video generation}
\subsection{Human Video Generation}
With advancements in diffusion models~\cite{ho2020denoising,song2020denoising}, significant progress has been made in image and video generation~\cite{nichol2021glide, ramesh2022hierarchical, li2023w-plus-adapter,mou2024t2i,huang2023composer,zhang2023adding,liu2023survey,rombach2022high,chen2023videocrafter1,geyer2023tokenflow,he2022latent,ho2022imagen,khachatryan2023text2video,wei2025personalized}. 
Among these developments, there is growing interest in human-centric video generation.
Recent methods~\cite{wang2023disco, chang2023magicdance,feng2023dreamoving,xu2024magicanimate,hu2024animate,zhu2024champ,tan2024animate} primarily focus on generating the character movements in the foreground, typically guided by human pose representations like OpenPose~\cite{wei2016convolutional,simon2017hand,cao2017realtime} and DensePose~\cite{guler2018densepose}. 
For example,
Disco~\cite{wang2023disco} employs ControlNet~\cite{zhang2023adding} to drive character movements, marking a significant step in applying diffusion models to human video generation. 
Following this, Dreamoving~\cite{feng2023dreamoving} incorporates the motion module from AnimateDiff~\cite{guo2023animatediff} to improve temporal consistency and employs IP-Adaptor~\cite{ye2023ip} to maintain facial identity. 
Both MagicAnimate~\cite{xu2024magicanimate} and AnimateAnyone~\cite{hu2024animate} use ReferenceNet to preserve detailed features of the reference character. 
Additionally, Champ~\cite{zhu2024champ} utilizes 3D parameters from the SMPL~\cite{loper2023smpl} model as driving signals, allowing for finer control over hand and facial movements. 
To generalize on anthropomorphic characters, Animate-X~\cite{tan2024animate} introduces the Pose Indicator, which captures the character movement patterns from the driving video both implicitly and explicitly. 
Notably, these methods mainly focus on the foreground character movement, often leaving the background either static or with minimal, uncontrollable movement. 
This lack of background dynamism can make the video feel flat, leading to visual fatigue for viewers. 
In contrast, a dynamic background enhances the mood by matching the character’s movements, adding vibrancy to the scene. 
In this paper, we introduce a camera-controllable video background generation framework that creates a more immersive and engaging experience by dynamically adjusting the background movement for any given static scene image.

\subsection{Video Inpainting}

With the camera's zoom-out or lateral movements, it is necessary to predict the new content from the given scene image. 
Many works have addressed image and video inpainting~\cite{ranftl2020towards,saharia2022palette,xie2023smartbrush,yang2023paint,fan2023hierarchical,yu2023magvit,zhou2023propainter,xu2024tunnel,pan2024actanywhere}. Among them, MAGVIT~\cite{yu2023magvit}, a masked generative video transformer, uses a 3D tokenizer and masked video token modeling to inpaint specific masked regions in a video. ProPainter~\cite{zhou2023propainter} enhances video inpainting by integrating dual-domain propagation and a mask-guided sparse video transformer. Tunnel Try-on~\cite{xu2024tunnel} addresses the video try-on challenge by using a ``focus tunnel" to preserve clothing details. ActAnywhere~\cite{pan2024actanywhere} automates the creation of video backgrounds given foreground subjects and scene images. However, ActAnywhere~\cite{pan2024actanywhere} predicts the background with an implicit motion from the foreground, which can easily result in mismatches between the foreground and background motions. To address this issue, we propose the new camera-controllable video background generation task,
{where the camera poses explicitly guide the background motion}.

\subsection{Motion-Conditioned Video Generation}
The earlier methods~\cite{he2022latent,hong2022cogvideo,karras2023dreampose,ruan2023mm,zhang2023i2vgen,chen2023videocrafter1} propose to control the content and motion of videos through text embeddings. Such textual descriptions provide only a coarse control over the video's motion. Latter approaches~\cite{esser2023structure,yin2023dragnuwa,zhang2023controlvideo,wang2024videocomposer,guo2025sparsectrl} use depth maps or other condition information from videos as control signals. However, these signals are still coupled with scene information, which hinders achieving a proper decoupling of camera movement. AnimateDiff~\cite{guo2023animatediff} learns different types of camera movement through LoRA~\cite{hu2021lora} but is often influenced by the appearances present in the training data. MotionDirector~\cite{zhao2025motiondirector} decouples appearance and motion learning by a dual-path LoRA adapter for motion customization. However, LoRA-based methods make it inconvenient to fine-tune the model for different types of camera movement. MotionCtrl~\cite{wang2024motionctrl} utilizes the extrinsic matrix of the camera to control the cinematography of generated videos. Building on this, CameraCtrl~\cite{he2024cameractrl} and HumanVid~\cite{wang2024humanvid} introduce Plücker embeddings~\cite{sitzmann2021light}, associating camera poses and video pixels to achieve more sophisticated camera movement generation.
In contrast, we differentiate ourselves from them by introducing a new camera-controllable video background generation framework. Our method explicitly aligns background dynamics with camera poses, achieving a higher level of background interactivity and consistency with foreground movements, creating a more immersive viewing experience.

\begin{figure*}[t]
	\setlength{\abovecaptionskip}{8pt} 
	\setlength{\belowcaptionskip}{-12pt}
	\centering
	\includegraphics[width=.98\textwidth]{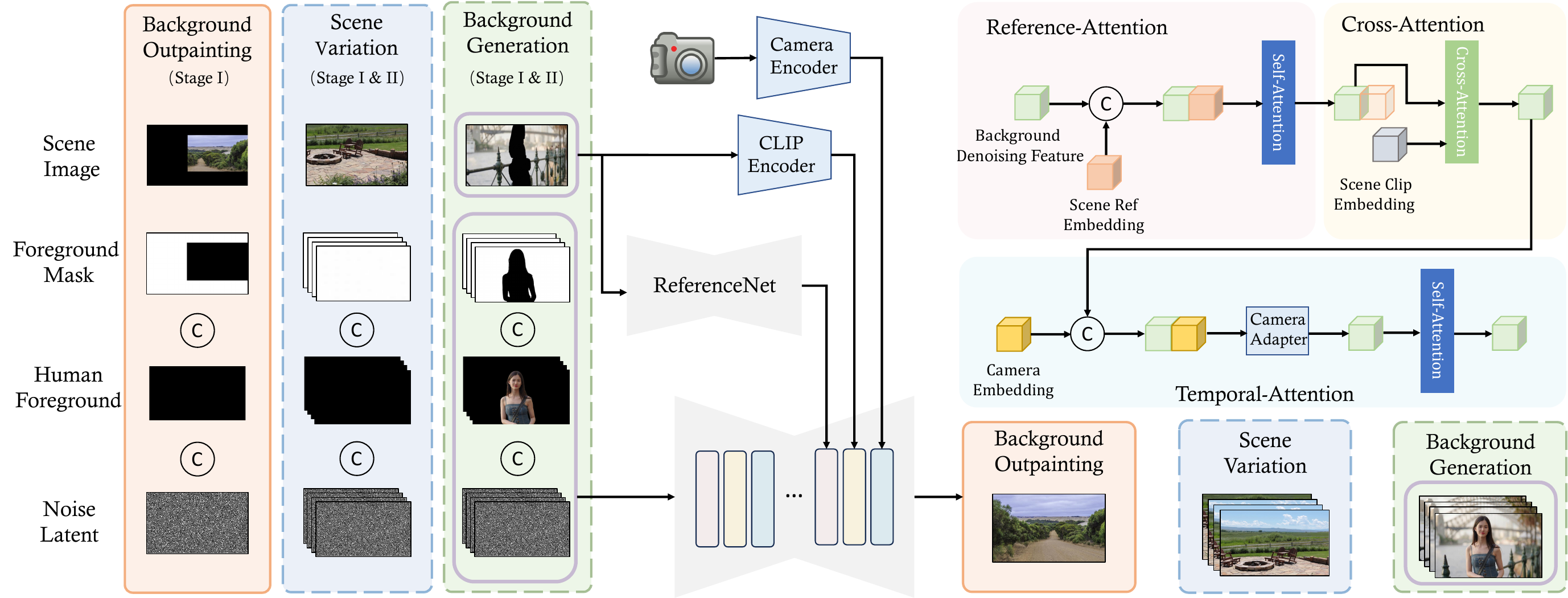}
	\caption{\textbf{Overview of DynaScene framework.} The noise latent, foreground mask, and human foreground are concatenated into the denoising U-Net. We employ the CLIP encoder and ReferenceNet to capture both high-level semantic features and fine-grained details from the scene image, respectively.  The camera pose is integrated into the Camera Encoder. To enhance the model's ability to generate coherent textures for newly revealed areas and preserve consistency in previously visible areas, we introduce multi-task learning including background outpainting in Stage I and scene variation across all stages. All tasks are trained on the same U-Net model.}
	\label{fig:method}
\end{figure*}

\section{Method}

As shown in Figure~\ref{fig:method}, our DynaScene framework takes three inputs, (1) a static scene image ${I}_s$ as the video background, (2) a sequence of human foreground frames $\{I_f^1,..., I_f^n\}$, and (3) the camera pose. 
The scene image $I_s$ is processed separately by a CLIP encoder and ReferenceNet, which provide the video generation model with both high-level semantic information and fine-grained textures through Cross- and Reference- Attention mechanisms.
The video foreground, foreground mask, and noise latent are concatenated together to the video Diffusion model. 
The camera pose controls the motion trend of the generated video. 
With these three inputs, the generated video background is expected to seamlessly integrate human motion into the scene image while also following the camera's movement.
This design has already demonstrated strong video generation capabilities in other tasks~\cite{wang2024humanvid,he2024cameractrl,wang2024motionctrl}.
To better capture the spatial relationship between the human foreground and the background and to ensure object consistency across multiple perspectives during camera movement, we propose a two-stage training approach focusing on spatial and temporal consistency. During each training stage, the entire network framework simultaneously trains on multiple tasks of background outpainting, scene perspective transformation, and background generation. 
During inference, only the background generation task is required.

\subsection{Preliminary of Latent Diffusion}
\label{sec:3.1}
Our model is based on Stable Diffusion 1.5 (SD~1.5)~\cite{rombach2022high}. As an extension of diffusion models~\cite{song2020denoising}, SD~1.5 utilizes a VAE~\cite{kingma2013auto} encoder $\mathcal{E}$ to compress the given image $\mathbf{x}_0$ into the latent space, $\mathbf{z}_0=\mathcal{E}(\mathbf{x}_0)$, which reduces computational complexity.
During the training process, diffusion models add Gaussian noise to $\mathbf{z}_0$ through Markov process~\cite{sohl2015deep} to obtain the noisy latent $\mathbf{z_t}$ at the corresponding time step $t$:
\begin{equation}
\small
\mathbf{z}_t=\sqrt{\bar{\alpha}_t} \mathbf{z}_0+\sqrt{1-\bar{\alpha}_t} \boldsymbol{\epsilon}, \boldsymbol{\epsilon} \sim \mathcal{U}([0,1])
\end{equation}
where $\bar{\alpha}_t$ represents the noise accumulation coefficient corresponding to $t$. Subsequently, the denoising network $\epsilon_\theta$ is employed to predict the added noise, with the objective defined as follows:
\begin{equation}
\small
\mathcal{L}=\mathbb{E}_{\mathbf{z}_t, c, \epsilon, t}\left(\left\|\epsilon-\epsilon_\theta\left(\mathbf{z}_t, c, t\right)\right\|_2^2\right)
\end{equation}
where $c$ is the text embeddings used for text-conditioned generation. During inference, we randomly sample noise $\mathbf{z}_T$ from a Gaussian distribution, and then iteratively denoise it over T time steps to obtain the final latent code $\hat{\mathbf{z}}_0$. Finally, $\hat{\mathbf{z}}_0$ is fed into the VAE decoder $\mathcal{D}$ to generate the final output: $\hat{\mathbf{x}}_0=\mathcal{D}(\hat{\mathbf{z}}_0)$.

\subsection{Two-Stage and Multi-Task Training Strategy}
\label{sec:3.2}
For the camera-controllable video background generation, there are several key challenges: 
1) with camera pose changes, newly revealed areas of the scene should exhibit coherent and realistic textures; 
2) as the camera view shifts, objects in the background should maintain visual consistency to preserve spatial coherence; and
3) the human subject should be realistically positioned within the background to avoid unnatural placements, such as floating mid-air.
To address the above issues, we propose a two-stage and multi-task training strategy. In Stage I: Image Background Generation, we enhance the realism and spatial consistency of newly generated background elements {with the foreground}. In Stage II: Video Background Generation, we focus on reinforcing consistency across consecutive frames to ensure temporal coherence in the video {and adding camera control to the generated background}. 

\noindent\textbf{Stage I: Image Background Generation.} 
To improve the continuity of video generation, we first focus on image-level training to enhance the model’s ability to generate realistic scenes. 
We observe that when the camera poses change, newly revealed areas must maintain realism and harmony, while previously visible areas should remain consistent. 
To address these challenges, we propose a multi-task learning framework. 
Specifically, we introduce a \textbf{background outpainting task} to improve visual consistency in newly revealed areas. 
Additionally, we propose a \textbf{scene variation task} to preserve the original textures when the camera poses shift. 
The background outpainting, scene variation, and image background generation tasks share the same diffusion model but with different inputs.

1) For the \textbf{background outpainting task}, we randomly mask 20\%$\sim$50\% of a random video frame as the scene image and obtain the corresponding foreground mask. The scene image is then fed into ReferenceNet and the CLIP encoder for further processing.
The human image foreground is set to 0, and the foreground mask, with a value of 1, indicates the region that is newly revealed and needs to be completed (see the 1-\textit{st} column in Figure~\ref{fig:method}). 

2) In the \textbf{scene image variation task}, the scene image is selected from video clips. The human foreground is filled with 0, while the foreground mask is set to 1. 
This design facilitates the generation of new viewpoints, which supports the consistency of new scenes for the subsequent stage of camera pose variation (see the 2-\textit{nd} column in Figure~\ref{fig:method}).

3) In the \textbf{image background generation task}, the image foreground is randomly selected from the video clip, while the background is taken from a different frame of the same video (see the 3-\textit{rd} column in Figure~\ref{fig:method}). 
Furthermore, we mask the human foreground in the image background and apply a dilation operation to alleviate the negative impact of the human appearance to form the scene image. This task enables the model to learn to adjust the scene image based on the human position, ensuring the correct placement of the foreground within the generated background and avoiding unnatural placements. 

In Stage I, we focus on the spatial domain and do not use the temporal module. 
These three tasks are trained within the same framework with equal probability. More results of background outpainting and image background generation are provided in the supplementary materials.

\noindent\textbf{Stage II: Video Background Generation.}
For video background generation, we integrate the motion module from AnimateDiff~\cite{guo2023animatediff} into the diffusion U-Net. The camera pose is incorporated into the motion module to model camera movements. In this stage, only the temporal attention is fine-tuned, while the other modules remain fixed.
Camera pose typically refers to the intrinsic parameters $\mathbf{K} \in \mathbb{R}^{3 \times 3}$ and extrinsic parameters $\mathbf{E} \in \mathbb{R}^{3 \times 4}$ of the camera. 
In this paper, we follow~\cite{he2024cameractrl} to adopt Plücker embedding for camera pose.
With the integration of camera pose, Stage II has three tasks trained simultaneously, \ie, scene video variation, video background generation, and adaptive background illumination adjustment.

1) The task of \textbf{scene video variation} is similar to that of scene image variation in Stage I, as it preserves the original textures when the camera poses shift. In Stage II, this task is designed to generalize to the temporal space of videos. Since the focus is primarily on background consistency rather than the human subject, it can be trained on other datasets, such as the Realestate10K dataset~\cite{zhou2018stereo}. In this stage, the scene image remains the same as in Stage I, but the foreground image and background mask are now a consecutive sequence from a video clip.

2) In the \textbf{video background generation task}, the model generates the video background by considering both the scene image and the human video foreground. This training primarily aims at the final application scenario, where the goal is to ensure that the generated video background not only aligns well with the human movement but also remains consistent with the given camera pose. By adjusting the scene image based on the foreground human’s position and the camera’s motion, the model learns to generate a realistic and coherent video background that fits seamlessly with both the human subject and the camera dynamics.

\lxm{3) The foreground human video and given scene image may have inconsistent lighting during inference. To address this issue, we introduce \textbf{adaptive background illumination adjustment} by applying random lighting augmentations to scene images during training. This enables the model to adjust the background lighting based on the foreground's illumination, ensuring a more natural integration.}

\subsection{Training Details}
\label{sec:3.4}

Our training process consists of two stages, designed to learn from both image and video domains. Before training, we initialize our denoising U-Net with the inpainting model weights~\cite{rombach2022high} from Stable Diffusion 1.5 and the ReferenceNet with the same weights. For the CLIP encoder, we use the CLIP-vit-large-patch-14 model.
In Stage I, we unfreeze the weights of the denoising U-Net and the ReferenceNet, while keeping the VAE and CLIP encoder fixed. During this stage, we randomly select a foreground frame and a background frame from the videos. In Stage II, we initialize the motion module with weights from AnimateDiff v2~\cite{guo2023animatediff} and train only the motion module.

\section{DynaScene Dataset}
\label{AnyScene Dataset}
A vivid video typically requires rich and dynamic movement. 
Currently, most human datasets~\cite{zhou2018stereo,zablotskaia2019dwnet,black2023bedlam,lin2024motion,wang2024humanvid}  consist of human videos where only the characters in the foreground are in motion, while the background scenery remains static or shows only minimal movement.
On the other hand, the Realestate10k dataset~\cite{zhou2018stereo}  includes videos capturing real estate scenes with dynamic camera movements, but it lacks any foreground characters. The recently released high-quality Humanvid dataset~\cite{wang2024humanvid} combines real-world (20k) and synthetic (75k) data for human image animation. However, we found that some of these videos either contain static scenes or feature minimal movement.
To better boost the human animation task, we propose the DynaScene dataset, which incorporates more dynamic camera movements and richer scene information. The DynaScene dataset consists of video foregrounds, scene images, and their corresponding camera poses.
A detailed comparison between the DynaScene dataset and existing datasets is provided in Table~\ref{tab:datasetcompare}. 
\lxm{Notably, our DynaScene dataset is 10 times larger than existing real-world video datasets.}
Our dataset also offers a notable advantage in resolution, and most importantly, the background movement is more pronounced compared to other methods. 
We believe that this dataset will provide excellent training data for human animation, enabling the generation of more vivid and dynamic videos. \lxm{To support future research, we will publicly release all video source links, annotations, and processing code, ensuring full reproducibility while respecting content ownership and copyright.}
Next, we detail the production process of the DynaScene dataset.

\subsection{Data Pre-processing} 

\lxm{The human videos are collected from short video platforms that allow downloads for non-commercial use.} These videos cover
a wide range of topics, such as cinematography tutorials, skiing, running, parkour, dancing, skateboarding, and roller skating. Many of these videos have dynamic camera movements, resulting in a total of 1,000 hours of engaging content.
We then filter out low-resolution videos, retaining only those with a resolution of 1080P or higher. To manage potential scene transitions, we follow the settings of the Scene Detect tool\footnote{https://github.com/Breakthrough/PySceneDetect} and segment transition scenes by calculating the difference in content between the current and previous frames. Finally, we use YOLO~\cite{redmon2016you} for human detection, removing videos that contain more than one person or lack any foreground characters.

\begin{table}[t]
\label{table:dataset}
	\centering
	\vspace{-0pt}
	\renewcommand\arraystretch{1.2}
        \footnotesize
	\setlength{\abovecaptionskip}{4pt}
	\setlength{\belowcaptionskip}{3pt}
	\caption{{Comparison with other human video datasets.}}
	\setlength{\tabcolsep}{0.4mm}
	{
		\begin{tabular}{l| c c c c c}
			\hline
   {\textbf{Datasets}}& Clips &  Size & CameraPose & DataType &Motion \\
			\hline
			TikTok~\cite{jafarian2021learning}      & 340 & $604\!\times\!1080$ & Static & Real &-  \\
			UBC-Fashion~\cite{zablotskaia2019dwnet} & 500 & $720\!\times\!964$ & Static & Real & -\\
			IDEA-400~\cite{lin2024motion}           & 12k & 720P & Static & Real& -   \\
			Bedlam~\cite{black2023bedlam}           & 10k & 720P & Dynamic & Syn.& 0.77   \\
			Humanvid~\cite{wang2024humanvid}        & 20\&75k & 1080P & Dynamic & Real\&Syn. & 0.74  \\
			\textbf{Ours}                           & \bf{200k} & \bf{1080P} & \bf{Dynamic} & \bf{Real} & \bf{0.64}  \\
			\hline
 \end{tabular}}
	\label{tab:datasetcompare}
	\vspace{-12pt}
\end{table}
\label{sec:4}

\subsection{Motion Intensity-Based Video Filtering} 
To create a human video dataset with richer and higher motion intensity, we further filter out videos with low motion intensity from the collected set.
Specifically, we follow the approach of LivePhoto~\cite{chen2023livephoto} and use the SSIM~\cite{ssim} score to define motion intensity.
A lower SSIM score indicates that the scenes are less similar, reflecting a higher degree of motion change.
In our process, we first extract frames from the video at a rate of 8 frames per second. Then, we calculate the SSIM score between every four consecutive frames. Finally, for a video clip with $n$ frames, the motion intensity score 
$S$ is obtained by averaging the SSIM values:
\begin{equation}
\small
\mathcal{S}=\sum_{i=0,i=i+4}^{n-4}{SSIM({I}_i,{I}_{i+4})}
\end{equation}
where $I_i$ represents the $i$-th frame of the video clip. 

To set an appropriate threshold for filtering out videos with low motion intensity, we manually select 200 video clips that exhibit rich movement based on human perception. We observe that over 80\% of these videos have a motion intensity score below 0.8. Therefore, we set the threshold for $S$ to 0.8: videos with $S \geq 0.8$ are classified as low-motion and excluded, while videos with $S < 0.8$ are considered to have richer motion and are retained in the final DynaScene dataset. 
The average motion intensity score $S$ of our final DynaScene dataset is 0.64, which significantly improves upon Humanvid~\cite{wang2024humanvid} (0.64 vs. 0.74).
Additionally, more than 37\% of the clips in Humanvid~\cite{wang2024humanvid} exhibit minimal camera movement ($S \geq 0.8$). This comparison highlights that our DynaScene dataset has a more diverse and richer motion, as detailed in  Table~\ref{tab:datasetcompare}.

\subsection{Video Foreground and Camera Pose Processing} 

We use a human-matting algorithm~\cite{liu2020boosting} to obtain the character’s foreground video. 
The reference scene image for each video clip is obtained by randomly sampling a frame from the video background.
For camera pose prediction, we employ the state-of-the-art VGGSFM algorithm~\cite{wang2024vggsfm}, which estimates camera poses by matching key points across frames. 
\lxm{Since our foreground characters are mostly in motion, we follow Humanvid~\cite{wang2024humanvid} and optimize VGGSFM by applying a human mask to exclude foreground key points, using only background key points for pose estimation. } This design significantly enhances pose accuracy. 
Finally, 50 qualified participants manually verify each predicted pose to ensure alignment with the video. As a result, the dataset is reduced by nearly half, resulting in a final DynaScene dataset of 200K clips.

\begin{table*}[tb!]
	\centering
        \vspace{-6pt}
	\renewcommand\arraystretch{1.05}
	\small
	\setlength{\abovecaptionskip}{4pt}
	\setlength{\belowcaptionskip}{3pt}
	\caption{{Quantitative comparison with competing methods. $\dagger$ indicates this method is re-implemented by us.}}
	\setlength{\tabcolsep}{3.56mm}
	{
		\begin{tabular}{l| c c c c c c c }
			\hline
   {\textbf{Methods}}& L1$\downarrow$ &  PSNR$\uparrow$ & SSIM$\uparrow$ & LPIPS$\downarrow$ & FID$\downarrow$ & FVD$_\text{I3D}\downarrow$  & {FVD}$_\text{3DRes}$$\downarrow$ \\
			\hline
			MotionCtrl~\cite{wang2024motionctrl} & 1.60e-04 &  28.13 & 0.382 & 0.596 &  285.16& 200.24 & 2780.03\\
			CameraCtrl~\cite{he2024cameractrl} & \underline{9.87e-05}  & \underline{29.12} & \underline{0.464} & \underline{0.442} & \underline{175.33} & \underline{104.11} & \underline{1770.60} \\
			ActAnywhere$^\dagger$~\cite{pan2024actanywhere}    & 1.42e-04 & 28.04 & 0.393 & 0.531 &  214.67 & 192.76 & 2174.41 \\
			\textbf{Ours}& \textbf{8.12e-05} & \textbf{29.27} & \textbf{0.506} & \textbf{0.354} & \textbf{96.18} & \textbf{55.84} & \textbf{1064.36} \\
			\hline
        \end{tabular}}
        \label{table:comparison}
	\vspace{0pt}
\end{table*}

\begin{figure*}[tb!]
	\setlength{\abovecaptionskip}{5pt} 
	\setlength{\belowcaptionskip}{-12pt}
	\centering
        \hspace{-10pt}
	\includegraphics[width=.91\textwidth]{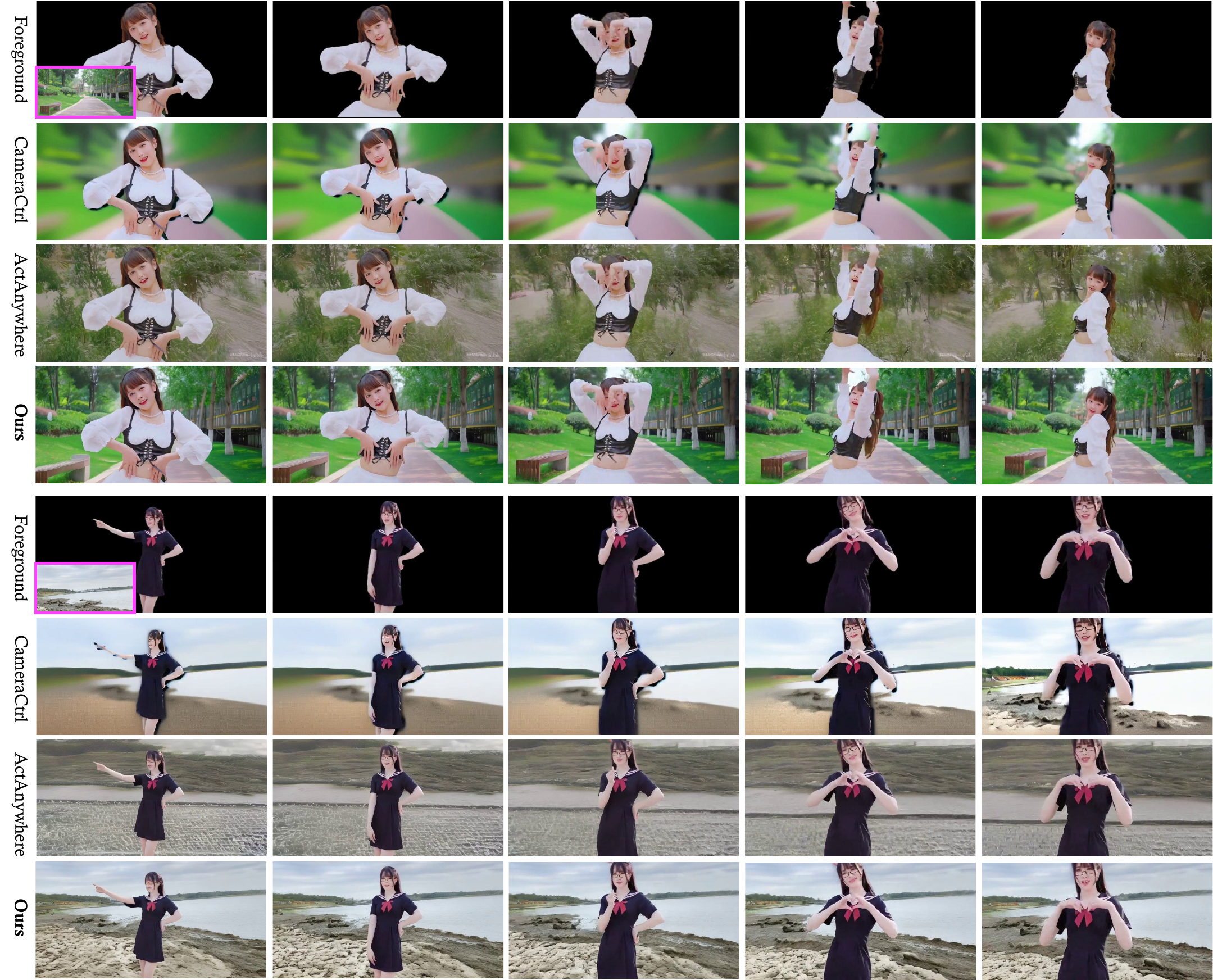}
	\caption{{Comparison with other methods.} The first and fifth rows are the video foreground, with the scene image located in the bottom-left corner of the first frame. The 2$\sim$4 and 6$\sim$8 rows are the results of CameraCtrl~\cite{he2024cameractrl}, ActAnywhere~\cite{pan2024actanywhere}, and our DynaScene, respectively.}
	\label{fig:comparsion}
\end{figure*}

\section{Experiments}
\noindent \textbf{Datasets}. Our proposed dataset consists of 204,957 video clips, split into three sets: 200,000 clips for training, 3,957 clips for validation, and 1,000 clips for testing, with no overlap between them. During training, we incorporate the Realestate10K dataset~\cite{zhou2018stereo} for the tasks of background outpainting and scene variation, while our DynaScene dataset focuses on human image and background generation.

\noindent \textbf{Evaluation Metrics}. For quantitative evaluation, we~use several metrics, including L1, PSNR, SSIM, LPIPS, FID~\cite{heusel2017gans}, and FVD~\cite{dong2019fw}, to assess both visual fidelity and perceptual quality. For the FVD evaluation, we employ two backbone models, \ie, I3D~\cite{carreira2017quo} and 3DRes~\cite{hara2018can}.

\noindent \textbf{Implementation Details.} During the whole training process, all images and video frames are resized to a resolution of $768 \times 432$. We use the Adam optimizer~\cite{kingma2014adam} with a learning rate of $1e{-5}$. 
All experiments are conducted on a server with 8 A100 GPUs. 
In Stage I, the images are randomly sampled from the video clip with a batch size of 3 and trained for 50,000 iterations. 
In Stage II, we sample 24 consecutive video frames with a batch size of 1 and train for 1,000,000 iterations. During inference, we use a DDIM sampler with 50 denoising steps.

\subsection{Comparison with Existing Methods}
Unlike these camera-controllable video generation methods, our approach mainly focuses on the realism of video background generation while preserving the human foreground movement. To validate its effectiveness, we compare our method with representative camera-controllable image-to-video generation methods (\eg, MotionCtrl~\cite{wang2024motionctrl}, CameraCtrl~\cite{he2024cameractrl}) and a video background generation method (\eg, ActAnywhere~\cite{he2024cameractrl}). 
Since ActAnywhere is not open source, we re-implement this method based on the details provided in their paper, 
\lxm{using the same training data as ours.}
For MotionCtrl and CameraCtrl, we use their SVD~\cite{blattmann2023stable} versions. During the denoising process of CameraCtrl and MotionCtrl at timestep $t$, we reintroduce the foreground latent $h_\textit{fore}^{t}$ into the original latent $h_\textit{ori}^{t}$ by
\begin{equation}
\small
{h}^{t} = h_\textit{ori}^{t} \times {M} + h_\textit{fore}^{t} \times (1-{M}),
\end{equation}
where $M$ represents the human foreground mask.

\begin{table*}[!t]
	\centering
	\renewcommand\arraystretch{1.05}
	\small
	\setlength{\abovecaptionskip}{4pt}
	\setlength{\belowcaptionskip}{3pt}
        \vspace{-6pt}
	\caption{{Comparison of DynaScene with camera control ($CC$), background outpainting ($BO$), and scene variation ($SV$).}}
	\setlength{\tabcolsep}{4.5mm}
	{
		\begin{tabular}{l| c c c c c c c}
			\hline
   {\textbf{Methods}}& L1$\downarrow$ &  PSNR$\uparrow$ & SSIM$\uparrow$ & LPIPS$\downarrow$ & FID$\downarrow$ & {FVD}$_\text{I3D}$$\downarrow$ & {FVD}$_\text{3DRes}$$\downarrow$ \\
			\hline
			Baseline   & 8.51e-05 & 28.99 & 0.495 & 0.364 & 98.63 & 59.32 & 1211.89 \\
			+ $CC$         & 8.41e-05 & \underline{29.10} & 0.500 & \underline{0.359} & 97.84 & 58.09 & 1124.56\\
			+ $CC$ + $BO$     & \underline{8.34e-05} & 29.06 & \underline{0.503} & 0.360 & \underline{96.73} & \underline{56.73} & \underline{1097.15}\\
			+ $CC$ + $BO$ + $SV$ & \textbf{8.12e-05} & \textbf{29.27} & \textbf{0.506} & \textbf{0.354} & \textbf{96.18} & \textbf{55.84} & \textbf{1064.36} \\
			\hline	
 \end{tabular}}
	\label{tab:ablation}
	\vspace{-18pt}
\end{table*}

\noindent \textbf{Quantitative Evaluation.} 
Table~\ref{table:comparison} presents a quantitative comparison between our method and others. 
The results show that our DynaScene outperforms these competing methods across various metrics on both pixel-level and perception-level evaluation. 
Specifically, our method demonstrates an improvement of 17.7\% and 19.9\% over the second-best method on L1 and LPIPS metrics, respectively. 
Notably, FID, FVD$_\text{I3D}$, and FVD$_\text{3DRes}$ show significant improvements of 45.14\%, 46.36\%, and 39.88\%, respectively. 
These results highlight the superior fidelity and smoothness of our method in generating human video backgrounds, leading to outputs that better align with human perception.

\begin{figure}[t]
	\setlength{\abovecaptionskip}{5pt} 
	\setlength{\belowcaptionskip}{-14pt}
	\centering
	\vspace{5pt}
	\includegraphics[width=0.455\textwidth]{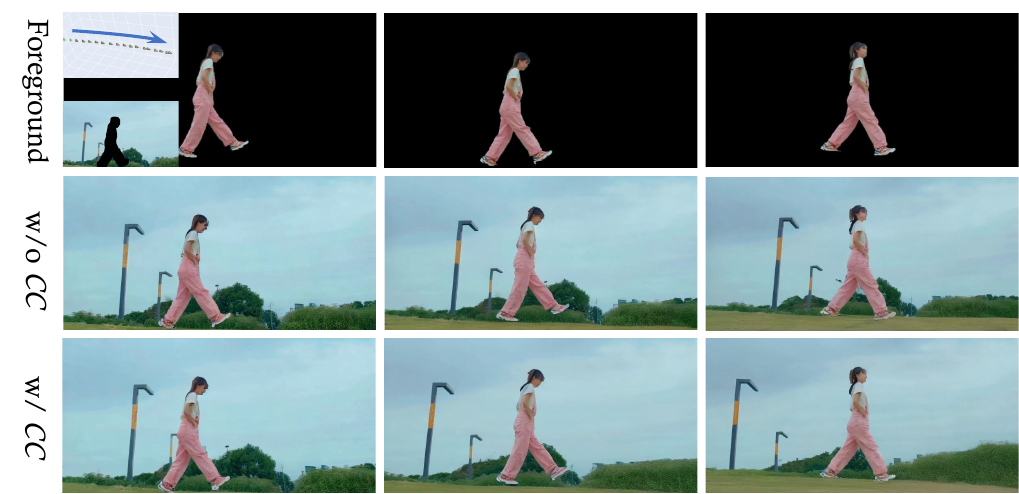}
	\caption{{Analyses of DynaScene w/ and w/o camera control ($CC$).} With $CC$, background aligns better with the foreground.}
	\label{fig:ablation+c}
        \vspace{-6pt}
\end{figure}

\begin{figure}[t]
	\setlength{\abovecaptionskip}{5pt} 
	\setlength{\belowcaptionskip}{-14pt}
	\centering
	\vspace{10pt}
	\includegraphics[width=0.455\textwidth]{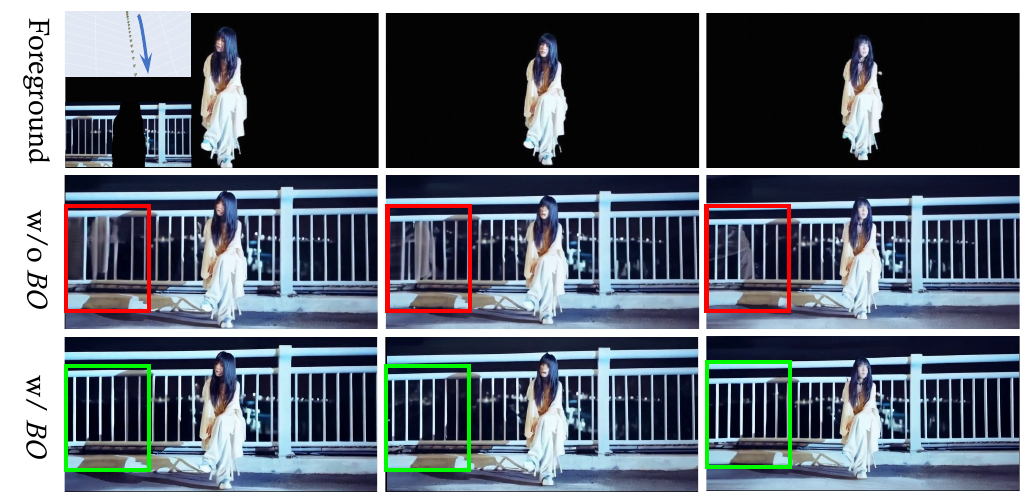}
	\caption{{Analyses of DynaScene w/ and w/o background outpainting ($BO$).} With $BO$, artifacts are removed effectively.}
	\label{fig:ablation+mo}
        \vspace{-4pt}
\end{figure}

\noindent\textbf{Qualitative Evaluation.} 
Figure~\ref{fig:comparsion} shows the visual comparison of our method with others. 
The results show that CameraCtrl~\cite{he2024cameractrl} can match the camera's movement with the foreground. However, when there is significant displacement between the generated background and the given scene image, the results become blurry and fail to preserve details effectively. ActAnywhere~\cite{he2024cameractrl}, a similar approach aimed at generating video backgrounds for human foregrounds, uses CLIP embeddings of the scene image as guidance for background generation. While this method captures high-level semantic information, it struggles to preserve finer details, leading to unsatisfactory results.
In contrast, our method leverages both ReferenceNet and CLIP embeddings to capture intricate details from the scene image. The two-stage, multi-task learning approach further enhances background consistency and realism. This enables the generation of video backgrounds that not only preserve human foreground movement with high fidelity but also align seamlessly with the scene image and camera pose. More results can be found in the supplemental materials.

\subsection{Ablation Study}
The main contributions of this paper include not only the newly introduced dataset but also the proposed multi-task training strategy. 
In this section, we evaluate the effectiveness of camera control and the multi-task training strategy, specifically focusing on background outpainting and scene variation. 
First, we train a baseline model without these components. Then, we gradually introduce camera control ($CC$), background outpainting ($BO$), and scene variation ($SV$) to the baseline model to validate their impact.

\noindent\textbf{Quantitative Evaluation.}
Table~\ref{tab:ablation} presents the evaluation metrics for our ablation experiments. 
Adding camera control ($CC$) contributes to an improvement across all metrics, with FVD$_\text{I3D}$ increasing by 7.7\% and FVD$_\text{3DRes}$ by 2.1\%, indicating enhanced temporal consistency and visual fidelity in video generation. 
Next, we integrate background outpainting ($BO$), leading to a 1.13\% improvement in FID, showing better real scene generation.
Finally, with the addition of scene variation~($SV$), our approach achieves optimal results in both pixel-based and perception-based metrics.

\begin{figure}[t]
	\setlength{\abovecaptionskip}{5pt} 
	\setlength{\belowcaptionskip}{-12pt}
	\centering
	\vspace{5pt}
	\includegraphics[width=0.46\textwidth]{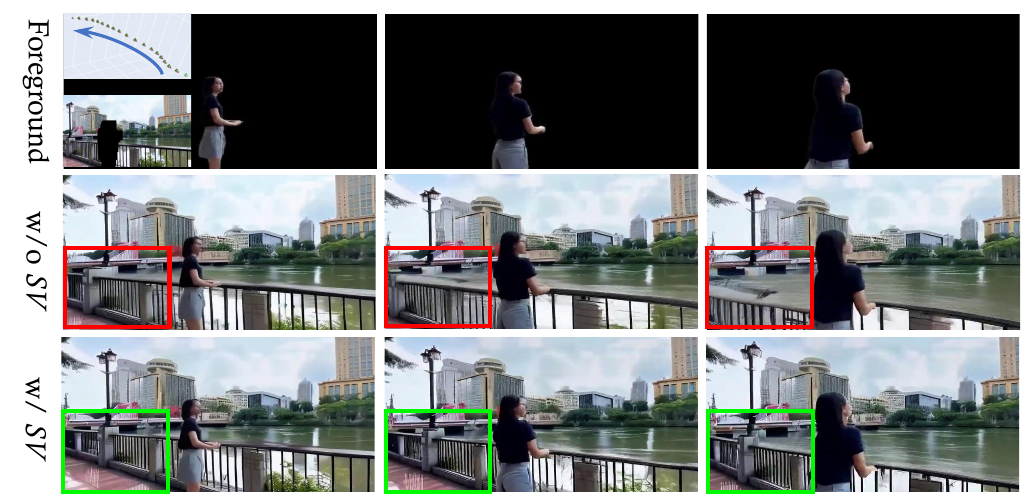}
	\caption{{Analyses of DynaScene w/ and w/o scene variation ($SV$).} With $SV$, the scenery are better-preserved ({\color{green}green} box).}
	\label{fig:ablation+mv}
    \vspace{-5pt}
\end{figure}

\noindent\textbf{Qualitative Evaluation.}
Figure~\ref{fig:ablation+c} compares results with and without camera control ($CC$). With the addition of camera control, the generated background’s motion pattern aligns more closely with the foreground, enhancing the continuity of the video background.
Figure~\ref{fig:ablation+mo} shows that introducing background outpainting ($BO$) allows the model to generate contextual elements beyond the scene image, reducing artifact occurrence. 
In Figure~\ref{fig:ablation+mv}, scene variation~($SV$) helps preserve the appearance of scene objects even as the camera’s viewpoint changes.
\lxm{These highlight the effect of our multi-task learning approach in this task. }

\section{Conclusion}

In this paper, we propose DynaScene, a new framework for camera-controllable video background generation. By incorporating camera movement control, DynaScene ensures motion consistency between the foreground and background. We also introduce a high-quality dataset, pairing video foregrounds, scene images, and camera poses, specifically designed for this task. To enhance performance, we employ a multi-task learning approach combining background outpainting and scene variation, enabling the generation of realistic and synchronized backgrounds.
DynaScene has broad applications in film production, virtual reality, and interactive gaming, enabling the creation of immersive, dynamic backgrounds that enrich user experiences.

{\small
\bibliographystyle{ieee_fullname}
\bibliography{egbib}
}
\clearpage
\appendix

\section{Camera Encoder and Camera Adaptor}\label{sec:A}

\begin{figure}[h]
	\centering
	\includegraphics[width=0.48\textwidth]{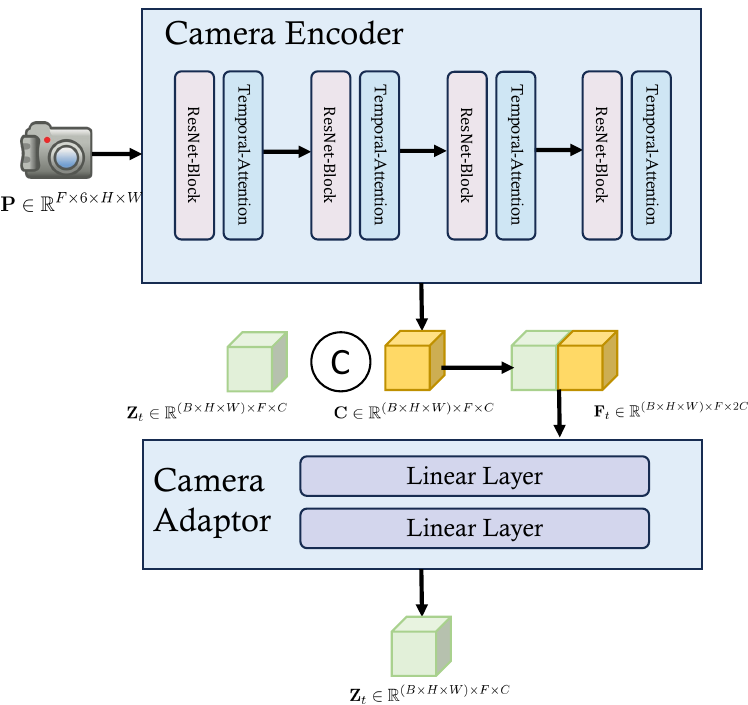}
	\caption{Details of camera encoder and camera adaptor.}
	\label{fig:CE}
\end{figure}

As illustrated in Figure~\ref{fig:CE}, the Camera Encoder consists of four blocks, each containing a ResNet Block and a Temporal-Attention module. Our Camera Encoder takes the Plücker embedding $\textbf{P} \in \mathbb{R}^{F \times 6\times H \times W}$ of the camera pose as input and generates multi-scale camera features $\textbf{C} \in \mathbb{R}^{(B \times H \times W) \times F \times C}$ for each block. Here, $B$, $H$, $W$, $F$, and $C$ represent the batch size, height, width, video length, and channel dimensions, respectively. These camera features are then concatenated with the background denoising features $\textbf{Z}_t \in \mathbb{R}^{(B \times H \times W) \times F \times C}$ before being processed by the Camera Adaptor. The Camera Adaptor employs two linear layers to effectively fuse the camera features with the denoising features. 
During the second stage of training, both the camera encoder and camera adapter are trainable.

\section{Pipeline of Processing DynaScene Dataset}
\label{sec:B}

\begin{figure}[h]
	\setlength{\belowcaptionskip}{-6pt}
	\centering
	\includegraphics[width=0.48\textwidth]{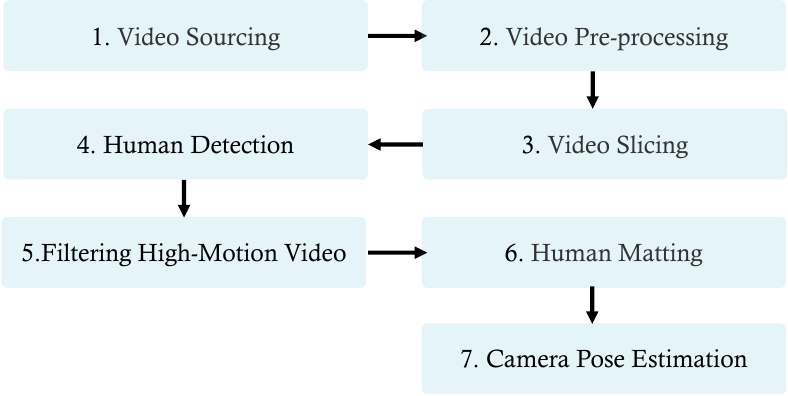}
	\caption{Pipeline of constructing DynaScene Dateset.}
	\label{fig:pipeline}
\end{figure}

Here, we briefly summarize the process of constructing the DynaScene Dataset. As shown in Figure~\ref{fig:pipeline}, 
the whole pipeline contains the following steps. 
1) Video Sourcing. We collect a large number of videos from open-source websites, focusing on dancing or sports themes, as they often exhibit high-intensity motion.
2) Video Pre-processing. Low-resolution videos are filtered out to ensure high-quality inputs. 3) Video Slicing. Videos are sliced into continuous segments, avoiding transitions between unrelated scenes.
4) Human Detection. We apply human detection~\cite{redmon2016you} to verify the presence of characters in the foreground and ensure that the proportion of the foreground is suitable throughout the video.
5) Filtering High-Motion Video. We evaluate the motion intensity of videos, retaining only those with significant motion. 6) Human Matting. A human matting algorithm~\cite{liu2020boosting} is employed to extract the human mask, human foreground, and scene image from each video frame. 7) Camera Pose Estimation. 
The camera pose of the background is estimated using VGGSFM~\cite{wang2024vggsfm}.

\section{Analyses of Two-stage Training Strategy}
\label{sec:C}

\begin{table}[h!]
	\centering
        \vspace{-9pt}
	\renewcommand\arraystretch{1.12}
        \scriptsize
	\setlength{\abovecaptionskip}{4pt}
	\setlength{\belowcaptionskip}{3pt}
	\caption{{Comparison of single- and two-stage training strategy.}}
	\setlength{\tabcolsep}{0.8mm}
	{
		\begin{tabular}{l| c c c c c c c}
			\hline
   {\textbf{Methods}}& L1$\downarrow$ &  PSNR$\uparrow$ & SSIM$\uparrow$ & LPIPS$\downarrow$ & FID$\downarrow$ & {FVD}$_\text{I3D}$$\downarrow$ & {FVD}$_\text{3DRes}$$\downarrow$ \\
			\hline
			single-stage   & 8.28e-05 &  29.09& 0.501 &  0.357& 98.44 & 57.54 & 1139.98   \\
			two-stage         & \textbf{8.12e-05} & \textbf{29.27} & \textbf{0.506} & \textbf{0.354} & \textbf{96.18} & \textbf{55.84} & \textbf{1064.36} \\
			\hline	
 \end{tabular}}
	\label{tab:trainint-strategy}
\end{table}

In our experiment, we implement a two-stage training strategy to enhance the video results. In the first stage, the goal is to generate human image backgrounds. Our DynaScene is trained on images to learn the spatial relationship between human foregrounds and scene backgrounds. 
In the second stage, we focus on generating human video backgrounds by integrating the motion module into our model and training it on videos. By using camera pose as a control signal for the background, 
we can generate a camera-controllable video background that maintains motion consistency with the human foreground. To evaluate the effectiveness of our two-stage training strategy, we conduct a comparison with a single-stage training strategy. 
In the single-stage training experiment, we train the model in the video domain using the same configuration as the second stage of the two-stage strategy, while unfreezing the weights of the ReferenceNet, Camera Encoder, Camera Adaptor, and denoising U-Net. 
The comparison between single-stage and two-stage training strategies is presented in Table~\ref{tab:trainint-strategy}. We can see that the two-stage training strategy leads to a significant reduction in FID and FVD scores, indicating a notable improvement in video continuity.

\section{More Results on Ablation Study}
\label{sec:D}

\begin{figure*}[t]
\renewcommand{\dblfloatpagefraction}{.9}

	\setlength{\abovecaptionskip}{5pt} 
	\setlength{\belowcaptionskip}{-14pt}
	\centering
	\vspace{10pt}
	\includegraphics[width=1\textwidth]{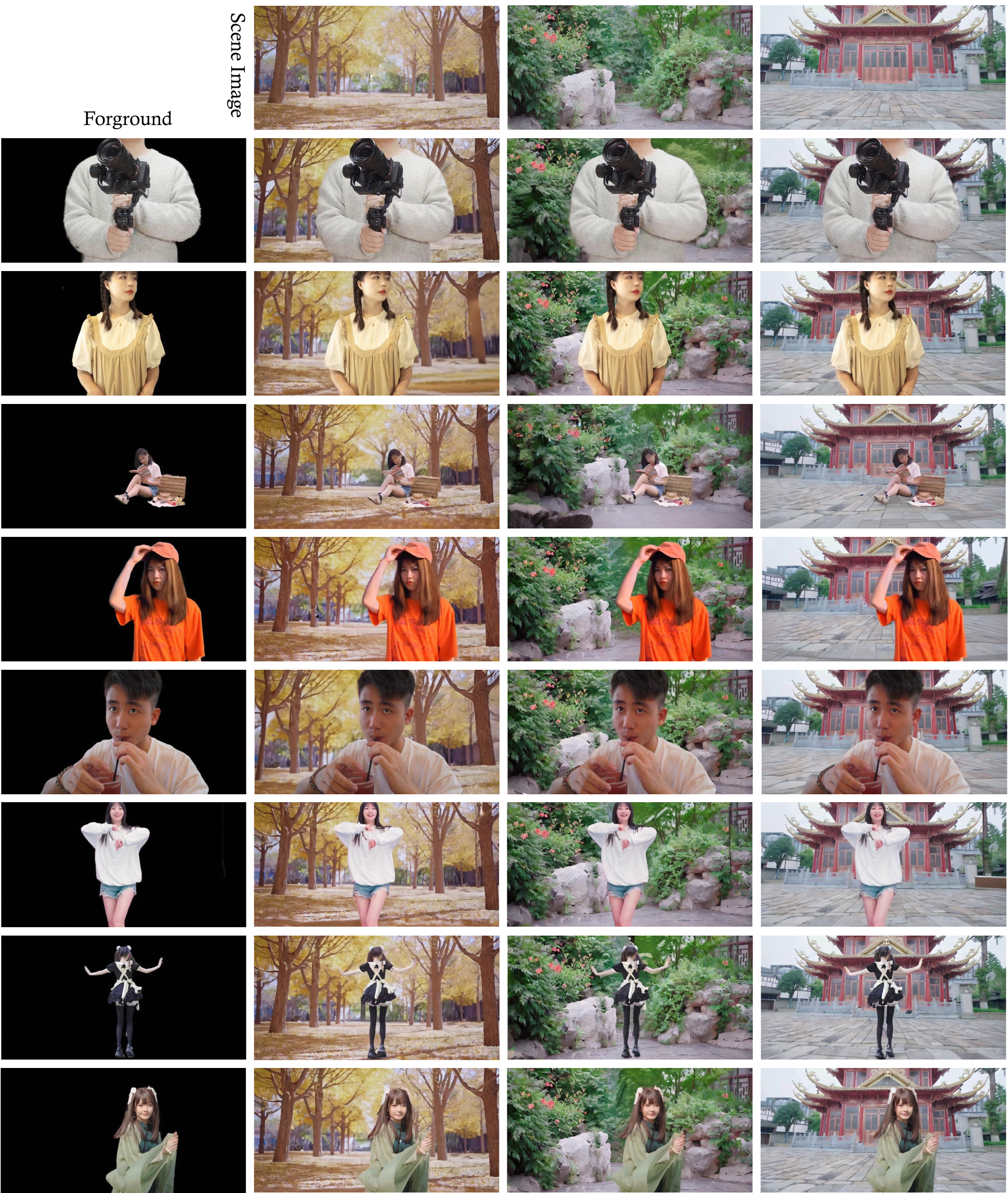}
	\caption{Results of human image background generation by DynaScene.}
	\label{fig:image_background}
\end{figure*}

\textbf{Human Image Background Generation.} Given a scene image and a human foreground image, DynaScene can effectively generate the corresponding background, ensuring the human is placed in the appropriate position within the scene (see Figure~\ref{fig:image_background}). 
This results in a natural and seamless integration of the human foreground into the background. 

\begin{figure*}[t]
	\setlength{\abovecaptionskip}{5pt} 
	\setlength{\belowcaptionskip}{-14pt}
	\centering
	\vspace{10pt}
	\includegraphics[width=1\textwidth]{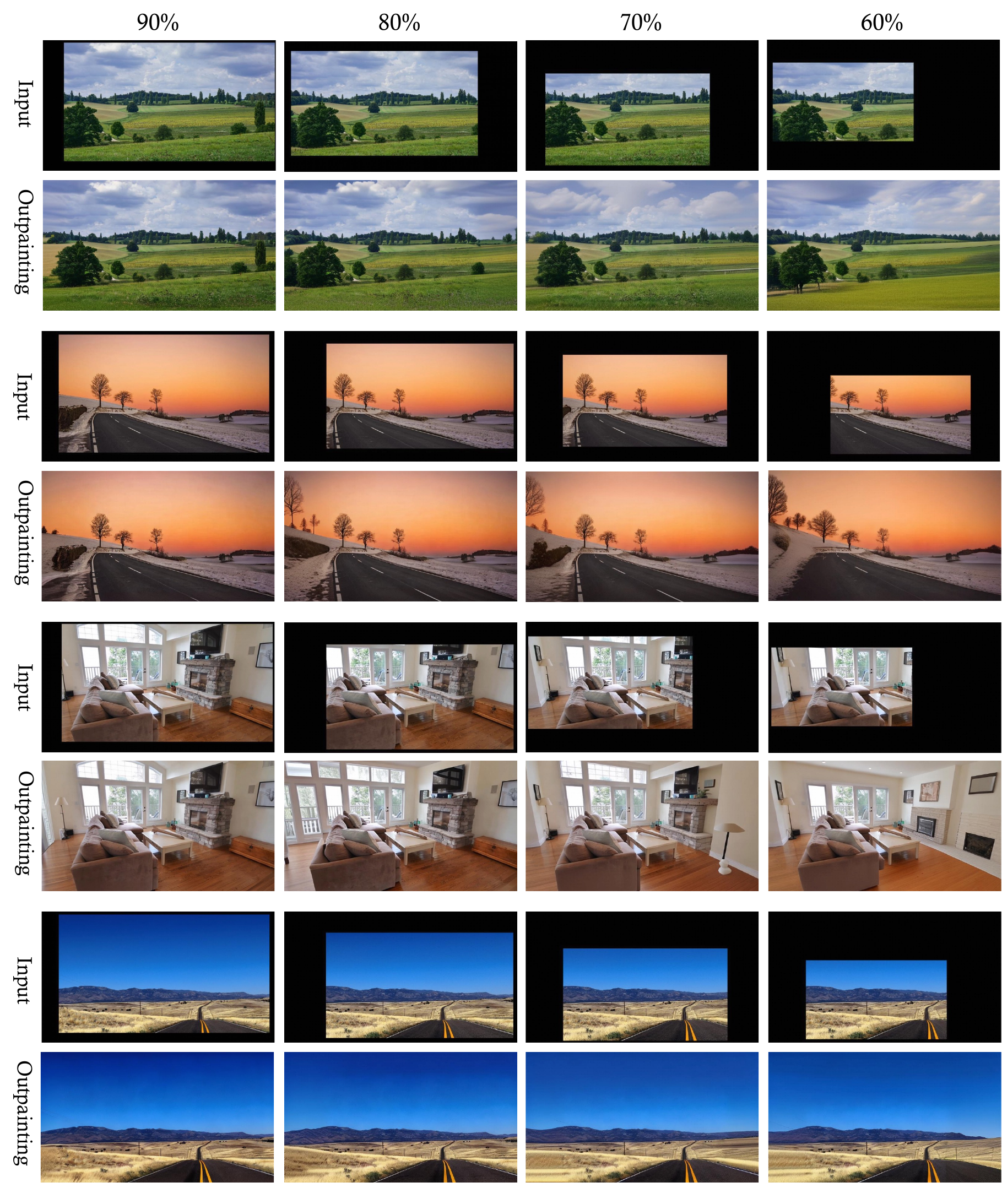}
	\caption{Background outpainting results by DynaScene. The original image is randomly masked with 90\%, 80\%, 70\%, and 60\% of the original image retained.}
	\label{fig:background_outpainting}
\end{figure*}

\noindent\textbf{Background Outpainting.} 
During the first training stage, we use background outpainting to enhance the model’s ability to generate realistic contextual elements in the background, particularly when the scene is viewed from new perspectives. This approach helps the model to more accurately generate background content for previously unseen viewpoints, improving the overall realism of the generated scenes. The visual results are shown in Figure~\ref{fig:background_outpainting}.

\begin{figure*}[t]
	\setlength{\abovecaptionskip}{5pt} 
	\setlength{\belowcaptionskip}{-14pt}
	\centering
	\vspace{10pt}
	\includegraphics[width=1\textwidth]{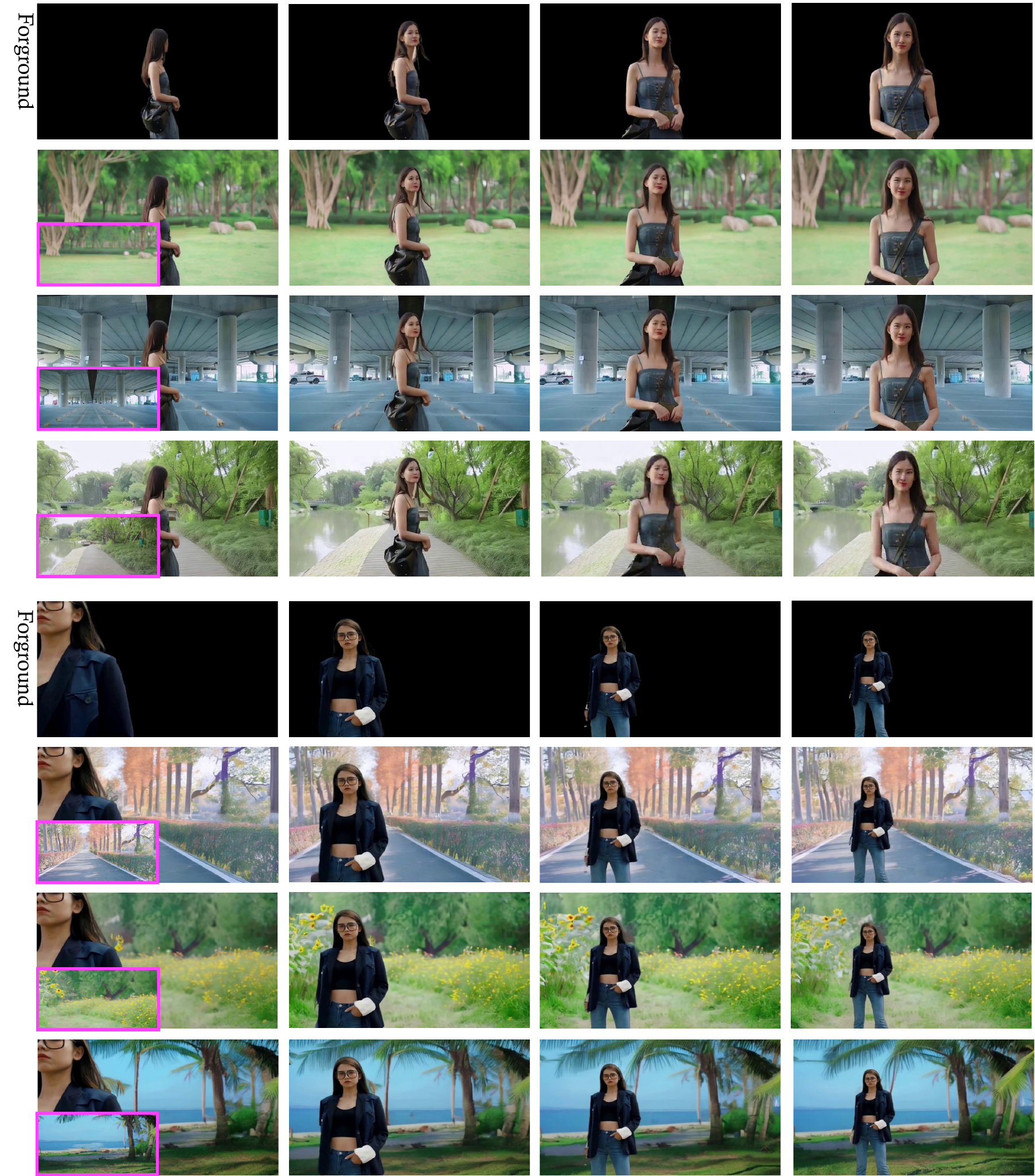}
	\caption{Results of human video background generation by DynaScene.}
	\label{fig:video_background}
\end{figure*}

\noindent\textbf{Human Video Background Generation.} 
As shown in Figure~\ref{fig:video_background}, we provide additional results for human video background generation. It is worth noting that our model not only supports scene images with human masks (black region) from DynaScene Dataset but also new scene images without masks. This demonstrates the strong generalization capabilities of our model, enabling it to generate backgrounds for a diverse range of scene images.

\begin{figure}[h]
	\centering
	\includegraphics[width=0.48\textwidth]{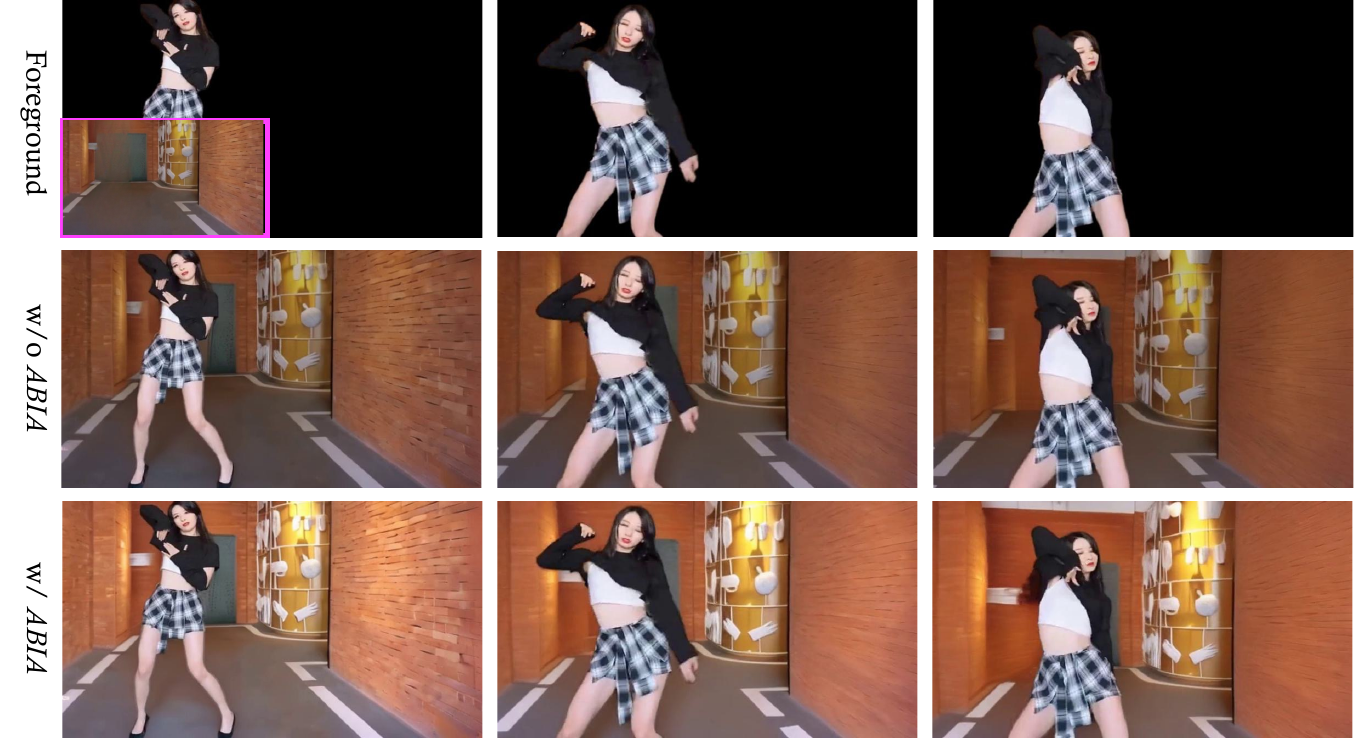}
	\caption{Analyses of DynaScene w/ and w/o adaptive background illumination adjustment ($\textit{ABIA}$). With $\textit{ABIA}$, the overall illumination is more consistent with the foreground human.}
	\label{fig:abia}
\end{figure}

\noindent\lxm{\textbf{Inconsistent Lighting during Inference.}
In practical applications, the illumination of a given scene image may differ from that of the foreground video. To ensure the illumination consistency between the generated background video and the foreground, we introduce a simple yet effective method: adaptive background illumination adjustment. 
Specifically, during training, we randomly modify the illumination of scene images while keeping the ground-truth video unchanged. This simulates real-world lighting inconsistencies between the scene and the human subject, enabling the model to learn to adaptively correct illumination mismatches. As shown in Figure~\ref{fig:abia}, we illustrate a case where the scene image has low illumination while the foreground subject has bright lighting. Without data augmentation, the generated background video retains the original scene image’s illumination, resulting in a noticeable inconsistency. In contrast, with adaptive background illumination adjustment, the model learns to generate a background video that harmonizes with the foreground’s lighting, ensuring a more natural and cohesive integration.
}

\begin{figure}[h]
	\centering
	\includegraphics[width=0.48\textwidth]{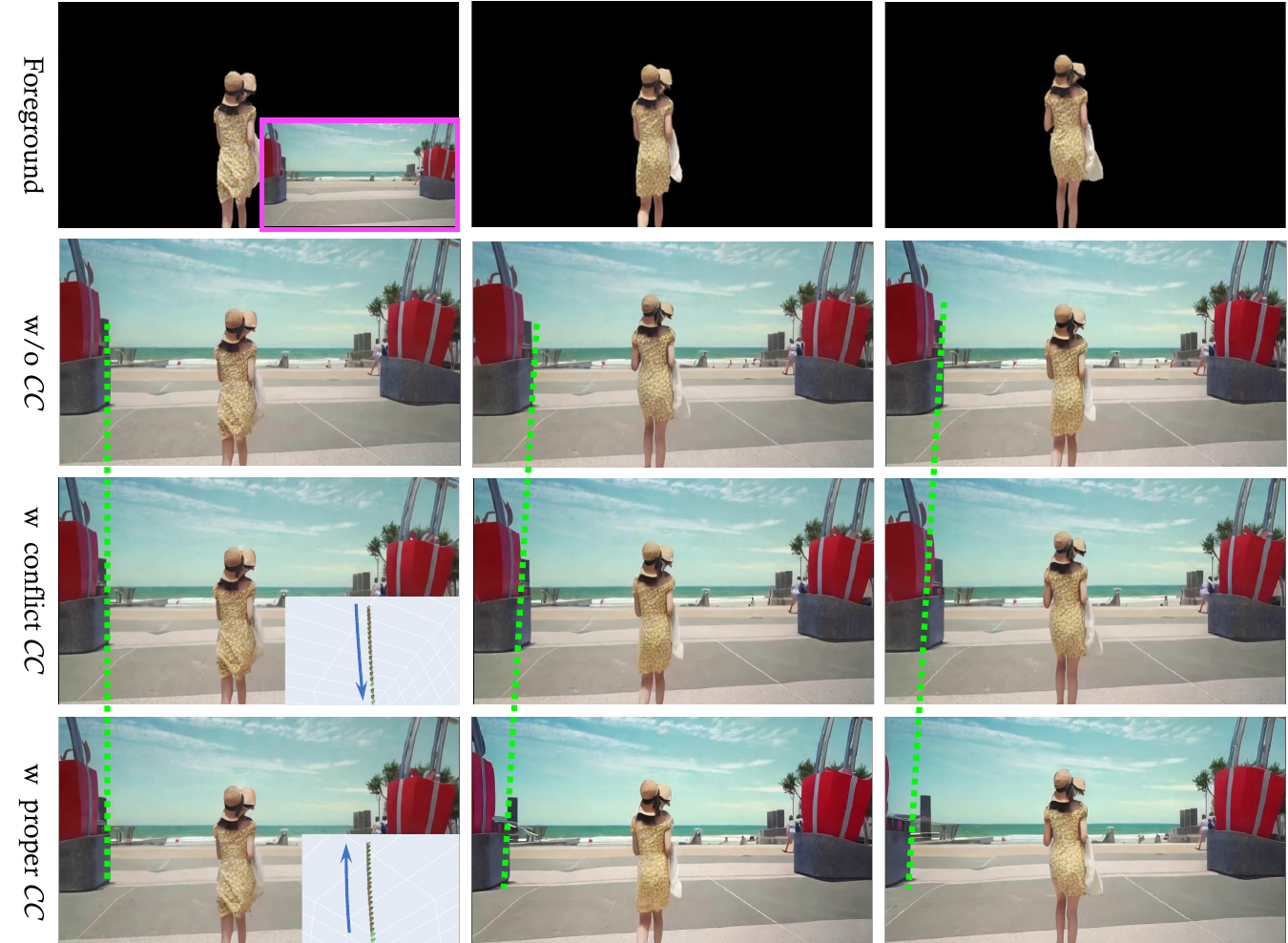}
	\caption{Conflicts between human foreground and camera pose.}
	\label{fig:motion}
\end{figure}

\noindent\textbf{\lxm{Analyses of Conflicts between Human and Camera Motion.}}
\lxm{Figure~\ref{fig:motion} illustrates a human foreground running forward. When no camera pose is given,  the generated background video remains largely static, making it appear as if the person is running in place. 
When an opposite camera pose is applied, the background exhibits subtle shaking, with its motion leaning more towards the given camera pose (see \textit{DynaScene.mp4}). In comparison, when a forward-moving camera pose is applied, which aligns with the foreground motion, the generated background accurately simulates forward movement, further validating the effectiveness of the explicit camera pose control.}

\section{Diversity Analyses of DynaScene Dataset}
\label{sec:E}
\begin{figure}[t]
	\setlength{\abovecaptionskip}{5pt} 
	\setlength{\belowcaptionskip}{-8pt}
	\centering
	\vspace{5pt}
	\includegraphics[width=0.48\textwidth]{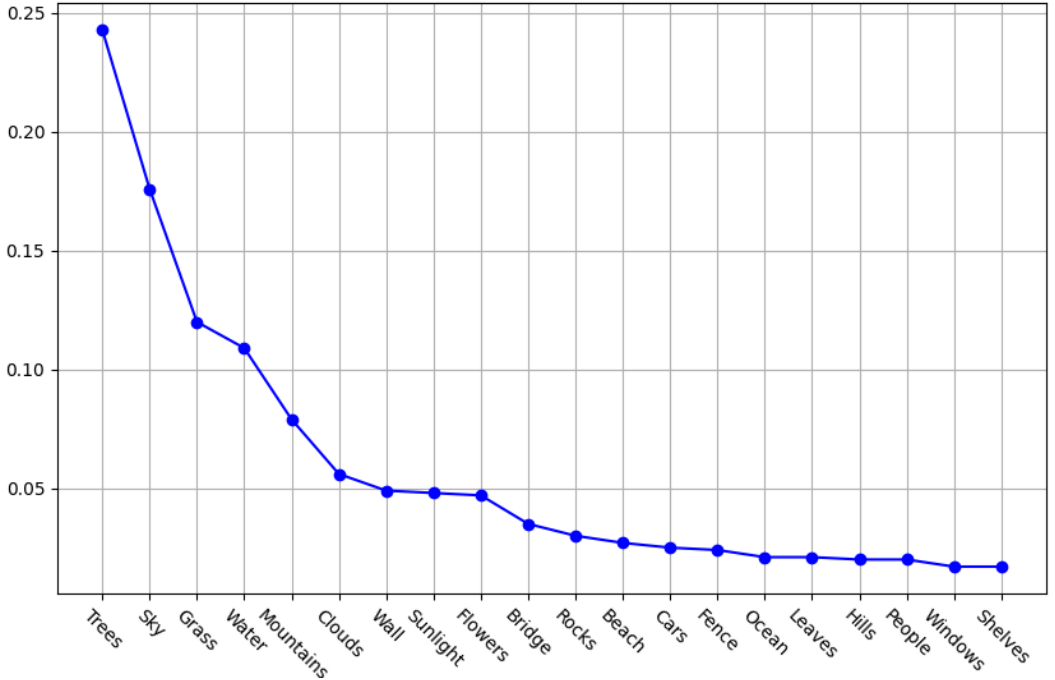}
	\caption{Scene element distribution in DynaScene dataset.}
	\label{fig:dataset_element}
\end{figure}

{The DynaScene dataset captures a diverse range of real-world environments, covering a wide variety of scene elements. To highlight its diversity, we employ LLAVA to identify the three primary background elements in each video. It reveals a broad spectrum of distinct scenes, including bridges, beaches, forests, and urban landscapes. As shown in Figure~\ref{fig:dataset_element}, we perform a statistical analysis of the 20 most frequently occurring elements. These elements include trees, sky, grass, water, mountains, clouds, wall, sunlight, flowers, bridges, rocks, beach, cars, fence, ocean, leaves, hills, people,  windows, shelves and so on, further highlighting the dataset’s diversity. This extensive variety of scene elements establishes a comprehensive and robust foundation for training experiments, enabling models to learn from diverse and realistic background conditions.}

\section{Limitations}
\label{sec:F}
\lxm{Our method may struggle to generate the desired results when given with conflicting inputs, such as a zoomed-in human foreground and a zoomed-out camera pose. 
Additionally, the quality of the extracted human foreground mask is critical to the overall performance. Inaccurate or noisy masks can result in unsatisfied foreground reconstruction, leading to noticeable artifacts or suboptimal blending at the edges between the foreground and the background.}

\end{document}


\title{Beyond Image Borders: Learning Feature Extrapolation for\\Unbounded Image Composition\\\begin{large}(Supplementary Material)\end{large}}

\author{Xiaoyu Liu$^1$, Ming Liu$^{1(}$\Envelope$^)$, Junyi Li$^1$, Shuai Liu, Xiaotao Wang, Lei Lei, Wangmeng Zuo$^{1,2}$ \\
$^1$Harbin Institute of Technology, Harbin, China \quad $^2$ Peng Cheng Laboratory, Shenzhen, China\\
{\tt\small 
\href{mailto:liuxiaoyu1104@gmail.com}{\color{black}liuxiaoyu1104@gmail.com}, \href{mailto:csmliu@outlook.com}{\color{black}csmliu@outlook.com}, \href{mailto:nagejacob@gmail.com}{\color{black}nagejacob@gmail.com}, \href{mailto:wmzuo@hit.edu.cn}{\color{black}wmzuo@hit.edu.cn}}
}

\maketitle
\ificcvfinal\thispagestyle{empty}\fi


The content of this supplementary material is organized as follows:

\begin{itemize}
    \item Difference between our proposed UNIC model and Zhong~\etal~\cite{zhong2021aesthetic} in \cref{sec:BIB_supp_introduction}.
    \item Details of the model architecture in \cref{sec:BIB_supp_model}.
    \item Details of the learning objective in \cref{sec:BIB_supp_learning_objective}.
    \item Details of our proposed datasets for unbounded image composition in \cref{sec:BIB_supp_dataset}.
    \item Details of the evaluation metrics in \cref{sec:BIB_supp_metrics}.
    \item Additional qualitative results \cref{sec:BIB_supp_results}.
\end{itemize}


\section{Difference with Zhong~\etal~\cite{zhong2021aesthetic}}
\label{sec:BIB_supp_introduction}
Since both our UNIC model and Zhong~\etal~\cite{zhong2021aesthetic} can give image composition results not fully lie in the camera view, we compare our UNIC with Zhong~\etal~\cite{zhong2021aesthetic} to show the difference.
%
The working schemes of the two methods are given in \cref{fig:NAIC_vs}.

\vspace{0.5em}
\noindent\textbf{Suitable Scenarios.}
As one can see, Zhong~\etal~\cite{zhong2021aesthetic} is designed to improve the composition of already taken images.
On the contrary, our approach provides recommendations to adjust the camera for obtaining a new camera view, which is suitable during the photography process.

\vspace{0.5em}
\noindent\textbf{Authenticity.}
Since Zhong~\etal~\cite{zhong2021aesthetic} extrapolate the given image via out-painting methods and crop in the extrapolated image, the results may contain unrealistic out-painted areas, which affects the image quality of the final results.
Our method instead guarantees that the final results are real images, and there are no concerns about authenticity.

\vspace{0.5em}
\noindent\textbf{Flexibility.}
Zhong~\etal~\cite{zhong2021aesthetic} highly rely on the already captured image, therefore the solution is greatly limited by the image out-painting methods.
Executing their method multiple times will lead to a certain result, leaving no chance to adjust the image composition by users.
Unlike their solution, our UNIC can be executed iteratively to obtain better results.
Besides, the users can get involved in the photography process naturally.

\begin{figure}[!t]
\centering
 \begin{overpic}[width=1.0\linewidth]{img/vs.pdf}
 \put(3,126){\footnotesize{Zhong~\etal~\cite{zhong2021aesthetic}}}
 \put(3,59){\footnotesize{UNIC}}
 \put(13,73){\scriptsize{Input Image}}
 \put(93,73){\scriptsize{Extrapolated Image}}
 \put(177,73){\scriptsize{Composition Result}}
 \put(90,121){\scriptsize{Not Real}}
 \put(156.5,108){\scriptsize{Image}}
 \put(152.5,97.5){\scriptsize{Cropping}}
 \put(60,108){\scriptsize{Image}}
 \put(54,97.5){\scriptsize{Outpainting}}
 \put(43,5){\scriptsize{Init View}}
 \put(152,5){\scriptsize{Recommended View}}
 \put(74,37){\scriptsize{Joint Camera Adjustment}}
 \put(79,24){\scriptsize{$ \& $ Image Composition}}
\end{overpic}
\vspace{1mm}
    \caption{Pipeline of Zhong~\etal~\cite{zhong2021aesthetic} and Our UNIC. The extrapolated areas by Zhong~\etal~\cite{zhong2021aesthetic} are not real, where obvious artifacts can be observed. By contrast, our UNIC jointly performs camera adjustment and image composition.}
    \label{fig:NAIC_vs}
\end{figure}

\section{Model Architecture}
\label{sec:BIB_supp_model}

Our UNIC contains a CNN backbone, a transformer encoder, a transformer decoder, and a feature extrapolation module.
The details of the architecture are given as follows.

\vspace{0.5em}
\noindent\textbf{CNN Backbone.}
Following \cite{meng2021conditional,carion2020end}, we use ResNet-50~\cite{he2016deep} as the backbone network.
For the initial image $\mathbf{I}_\mathit{init}\in\mathbb{R}^{3\times \mathit{H}_0\times \mathit{W}_0}$, we only use the last-layer feature $\mathbf{h}_\mathit{init}$ extracted by backbone, where $\mathbf{h}_\mathit{init}\in\mathbb{R}^{\mathit{C}\times \mathit{H}\times \mathit{W}}$, $\mathit{H}=\tfrac{\mathit{H}_0}{32}, \mathit{W}=\tfrac{\mathit{W}_0}{32}, \mathit{C} = 2048$. 

\vspace{0.5em}
\noindent\textbf{Transformer Encoder.}
We first employ a $1 \times 1$ convolution to reduce the number of channels of the feature from 2048 to 256.
The image features with encoded positional embeddings are rearranged into a sequence of feature tokens that can be fed into the encoder.
The encoder is composed of six standard transform blocks including a multi-head self-attention module and a feed-forward network (FFN), where the query and key are provided by different feature tokens.

\vspace{0.5em}
\noindent\textbf{Feature Extrapolation Module (FEM).}
The FEM is a stack of decoder layers to predict the padded features conditioned on visible ones.
The query is initialized by a learnable variable $\mathbf{m}$ and is transformed into padded features $\mathcal{Z}_\mathit{pad}$ by multiple decoder layers and a feed-forward network.
The detailed structure of the decoder layer of FEM is illustrated in \cref{fig:NAIC_tf_block}.
In each decoder layer, apart from self-attention between different padded feature embeddings, we also calculate cross-attention between visible features and padded features to utilize the visible information to learn feature extrapolation.
The key and value for cross attention consist of visible features $\mathcal{Z}_\mathit{vis}$ and the output of the self-attention layer.

\vspace{0.5em}
\noindent\textbf{Transformer Decoder.}
%
 A group of learnable anchors  are transformed into output embeddings by the decoder and are futher converted into bounding boxes and their corresponding confidence labels by two heads.
The decoder is standard architecture of the
transformer, which perform through multiple multihead self-attention between different decoder embeddings to remove redundant boxes and and cross attention aggregating image information to predict the boxes.

\begin{figure}[!t]
\centering
 \begin{overpic}[width=0.6\linewidth]{img/transformer_block.pdf}
 \put(0.5,36){\footnotesize{$\mathcal{Z}_\mathit{vis}$}}
  \put(34,189){\footnotesize{$\mathcal{Z}_\mathit{pad}$}}
    \put(38,4.5){\footnotesize{$\mathbf{m}$}}
\end{overpic}
\vspace{1mm}
    \caption{Illustrating the first layer of FEM. Later layers take the output of the previous layer as input.}
    \label{fig:NAIC_tf_block}
\end{figure}

\section{Learning Objective}
\label{sec:BIB_supp_learning_objective}

\begin{figure*}[!t]
\centering
 \begin{overpic}[width=0.92\linewidth]{img/example.pdf}

\end{overpic}
\vspace{1mm}
    \caption{Some sample images from our datasets. The left is the full-view image from the GAICD\cite{zeng2019reliable}, and the right is the image from our dataset. The ground-truth views are circled by the red box}
    \label{fig:NAIC_example}
\end{figure*}

\vspace{0.5em}
\noindent\textbf{Unbounded Regression on Set Prediction.}
Our network predicts a set of $\mathit{N}$ bounding boxes, but the number of predicted boxes is not consistent with the number of ground truth views $\mathit{N}^\mathit{gt}$ (typically $\mathit{N}^\mathit{gt} \ll \mathit{N}$).
Following \cite{meng2021conditional,carion2020end,jia2022rethinking}, we use set prediction to compute the loss in a reasonable way.
First, we pad the ground truth set of views with $\varnothing$ (invalid view) to a set of size $\mathit{N}$ as follows:
\begin{equation}
\begin{cases}
\left\{\mathbf{c}^\mathit{i}=\left[\mathit{x}, \mathit{y}, \mathit{w}, \mathit{h}\right], \mathbf{p}^\mathit{i}=1\right\}, & 1 \leq \mathit{i} \leq N^{\mathit{gt}} \\
\left\{\mathbf{c}^\mathit{i}=\varnothing, \mathbf{p}^\mathit{i}=0\right\}, & N^{\mathit{gt}}+1 \leq \mathit{i} \leq N 
\end{cases},
\end{equation}
then we use a bipartite matching to find an one-to-one index mapping $\sigma \in \mathfrak{S}_{N}$ for these two set to minimize the matching cost $\mathcal{L}_\mathrm{comp}$ which is the same as the loss function,\footnote{More intuitively, we can regard $\sigma$ as a shuffle operation, and $\mathfrak{S}_\mathit{N}$ is all shuffle solutions. We aim to find a shuffle solution $\sigma^\ast$ such that the loss $\mathcal{L}_\mathrm{comp}$ is minimized.}
 \begin{equation}
      \sigma^\ast=\underset{\sigma \in \mathfrak{S}_{N}}{\arg \min } \sum_{i}^{N} \mathcal{L}_{\mathrm {comp }}\left(\{\mathbf{c}^{\sigma(\mathit{i})}_\mathit{pred},\mathbf{p}^{\sigma(\mathit{i})}_\mathit{pred}\},\{\mathbf{c}^\mathit{i},\mathbf{p}^\mathit{i}\}\right).
 \end{equation}

Those predicted views having a matching with ground-truth valid views (\ie, $\mathbf{c}_{\mathit{i}} \neq \varnothing$) contribute to the regression loss, IoU loss, and focal loss (with $\mathbf{p}_\mathit{i}=1$).
For other views without a valid matching (\ie, $\mathbf{c}_{\mathit{i}} = \varnothing$), they only contribute to the focal loss (with $\mathbf{p}_\mathit{i}=0$).
With the above definitions, the loss function of our UNIC can be rewritten as
\begin{equation}
\begin{split}
    \mathcal{L}_\mathrm{comp} =\sum_{\mathit{i}=1}^\mathit{N} & \mathit{f}_\mathrm{bool}(\mathbf{c}_{\mathit{i}} \neq \varnothing) \mathcal{L}_\mathrm{reg}(\mathbf{c}_\mathit{pred}^{\sigma^\ast(\mathit{i})}, \mathbf{c}^{\mathit{i}}) + \\ & \mathit{f}_\mathrm{bool}(\mathbf{c}_{\mathit{i}} \neq \varnothing) \lambda_\mathrm{IoU}\mathcal{L}_\mathrm{IoU}(\mathbf{c}_\mathit{pred}^{\sigma^\ast(\mathit{i})}, \mathbf{c}^{\mathit{i}}) + \\ & \lambda_\mathrm{focal}\mathcal{L}_\mathrm{focal}(\mathbf{p}_\mathit{pred}^{\sigma^\ast(\mathit{i})}, \mathbf{p}^{\mathit{i}}),
\end{split}
\label{eqn:BIB_supp_loss1}
\end{equation}
where $\mathit{f}_\mathrm{bool}(\ast)$ equals to 1 when the condition $\ast$ is satisfied otherwise 0.

\vspace{0.5em}
\noindent\textbf{Smooth Label.}
It is worth noting that $\mathit{N}^\mathit{gt}$ is typically much smaller than $\mathit{N}$, therefore, the invalid views whose $\mathbf{p}_\mathit{i}=1$ have major contributions to $\mathcal{L}_\mathrm{focal}$ in \cref{eqn:BIB_supp_loss1}.
To remedy this problem, Jia~\etal~\cite{jia2022rethinking} have proposed two strategies to smooth the labels for invalid views, \ie, quality guidance and self-distillation.
The quality guidance strategy is more feasible for densely annotated datasets (\eg, GAICD~\cite{zeng2019reliable}), which uses the annotated quality score to get the smooth labels of invalid views.
For a predicted invalid view, we calculate the IoU between the view and all labeled views, and regard the quality score of maximum-IoU
neighbor view as the quality score of the predicted box.
Then we can map the quality score to the soft labels by a linear function.
As for the self-distillation strategy, the smooth labels are generated by the model whose parameters are exponential moving averages (EMA) of the model parameters.
%
%
Jia~\etal~\cite{jia2022rethinking} use these strategies for GAICD and CPC, respectively.

In this paper, we argue that, at the beginning of the training process, the model is not well trained, thus the self-distillation strategy is unable to generate stable smooth labels.
On the contrary, at the later phase of the training process, most predicted views have decent quality but are unable to match with limited ground-truths, forcing $\mathbf{p}_\mathit{i}$ to 0 or a manually set label is unreasonable.
Therefore, we propose to combine these two strategies based on our observations, and adopt the quality guidance strategy at first, then switch to the self-distillation strategy afterward.

\vspace{0.5em}
\noindent\textbf{Loss Terms.}
We employ the learning objective for composition including a regression loss $\mathcal{L}_\mathrm{reg}$ and a generalized IoU loss $\mathcal{L}_\mathrm{IoU}$ for bounding box regression, and a focal loss $\mathcal{L}_\mathrm{focal}$ for predicting the confidence.
\begin{equation}
\begin{split}
    \mathcal{L}_\mathrm{comp} &= \mathcal{L}_\mathrm{reg}(\mathbf{c}_\mathit{pred}, \mathbf{c}) + \lambda_\mathrm{IoU}\mathcal{L}_\mathrm{IoU}(\mathbf{c}_\mathit{pred}, \mathbf{c}) \\
    &+ \lambda_\mathrm{focal}\mathcal{L}_\mathrm{focal}(\mathbf{p}_\mathit{pred}, \mathbf{p}).
\end{split}
\end{equation}
The regression loss is defined as an $\ell_1$ loss to supervise $\mathbf{c}_\mathit{pred}$, \ie, a four-dimensional vector consisting of the box center coordinates and its height and width, \ie,
\begin{equation}
\begin{aligned}
    \mathcal{L}_\mathrm{reg}(\mathbf{c}_\mathit{pred}, \mathbf{c})=\|\mathbf{c}_\mathit{pred}-\mathbf{c}\|_1,
\end{aligned}
\end{equation}
while the IoU loss is defined by,
\begin{equation}
\begin{aligned}
    \mathcal{L}_\mathrm{IoU}(\mathbf{c}_\mathit{pred}, \mathbf{c})=1-\left(\frac{|\mathbf{c}_\mathit{pred} \cap \mathbf{c}|} {|\mathbf{c}_\mathit{pred} \cup \mathbf{c}|}-\frac{|\mathbf{c}_\mathit{c} - \mathbf{c}_\mathit{pred} \cup \mathbf{c} |}{|\mathbf{c}_\mathit{c}|}\right),
\end{aligned}
\end{equation}
where $|\cdot|$ means area, 
 and $\mathbf{c}_\mathit{c}$ is the smallest enclosing convex object for $\mathbf{c}_\mathit{pred}$ and $\mathbf{c}$.
And the focal loss is,
\begin{equation}
\begin{aligned}
    &\mathcal{L}_\mathrm{focal}(\mathbf{p}_\mathit{pred},\mathbf{p})=-|\mathbf{p}_\mathit{pred}-\mathbf{p}|^\beta \\
    & ((1-\mathbf{p}_\mathit{pred}) \log (1-\mathbf{p})+\mathbf{p}_\mathit{pred} \log (\mathbf{p})),
\end{aligned}
\end{equation}
where $\beta=2$.

\section{Unbounded Image Composition Dataset}
\label{sec:BIB_supp_dataset}
We construct the unbounded image composition (UIC) dataset based on the existing datasets (\ie, GAICD \cite{zeng2019reliable}, CPC \cite{wei2018good} \& FLMS \cite{fang2014automatic}).
In our UIC dataset, the ground-truth composition may not fully lie in the range of the image, and the rules for generating the dataset are given in the main manuscript.
Some examples of the UIC dataset are shown in \cref{fig:NAIC_example}.

\section{Evaluation Metrics}
\label{sec:BIB_supp_metrics}
In this section, we provide the formulations of the evaluation metrics used in our paper, \ie, $\mathit{Acc}_{\mathit{K}\!/\!\mathit{N}}$, IoU, and Disp.

$\mathit{Acc}_{\mathit{K}\!/\!\mathit{N}}$ is the accuracy (or ratio) of predicted $\mathit{K}$ views fall into the $\mathit{N}$ ground-truth views.
In specific, for image $\mathit{i}$, we define the set of annotated views with the top $\mathit{N}$ of quality score as $\mathit{C}_\mathit{N}^\mathit{i}=\left\{\mathbf{c}^{\mathit{i}1},\ldots,\mathbf{c}^{\mathit{ij}},\ldots, \mathbf{c}^{\mathit{iN}}\right\}$.
And a model returns $\mathit{K}$ views with the highest confidence scores denoted by $\mathbf{c}^{\mathit{ik}}_{\mathit{pred}}$.
Then $\mathit{Acc}_{\mathit{K}\!/\!\mathit{N}}$ can be defined by
\begin{equation}
    \mathit{Acc}_{\mathit{K}\!/\!\mathit{N}}=\frac{1}{\mathit{TK}} \sum_{i=1}^\mathit{T} \sum_{k=1}^\mathit{K} \mathit{f}_\mathrm{bool}(\max _{\mathbf{c}^{ij} \in {B}_N^i}\left\{{\rm IoU}\left(\mathbf{c}^\mathit{ik}_\mathit{pred}, \mathbf{c}^\mathit{ij}\right)\right\} \geq \epsilon),
\end{equation}
where $\epsilon\in\{0.90, 0.85\}$, $\mathit{f}_\mathrm{bool}(\ast)$ equals to 1 when the condition $\ast$ is satisfied otherwise 0.

The boundary displacement error (Disp.) is defined by
\begin{equation}
    \textrm{Disp} = \frac{1}{4}\sum_{\mathit{j}} \|\mathbf{b}_\mathit{pred}^{\mathit{j}}-\mathbf{b} ^{\mathit{j}}\|_1
\end{equation}
where $\mathbf{b}$ denotes all boundaries of the bounding box, and $\mathbf{b}_\mathit{j}$ is the normalized coordinate of that boundary (\eg, the x-axis of the left and right boundary).
Finally, the IoU is defined by, 
\begin{equation}
    \textrm{IoU}=\frac{|\mathbf{c}_\mathit{pred} \cap \mathbf{c}|} {|\mathbf{c}_\mathit{pred} \cup \mathbf{c}|}.
\end{equation}

\section{Additional qualitative results.}
\label{sec:BIB_supp_results}
In \cref{fig:NAIC_compare_flms}, we present the visualization comparison with other existing cropping methods on FLMS~\cite{fang2014automatic} dataset. 
%
%
Meanwhile, we give additional visualization results in \cref{fig:NAIC_compare} on GAICD ~\cite{zeng2019reliable} dataset.
And it can be seen that our method usually produces more appealing results by unbounded image composition.

\begin{figure*}[htp]
\centering
 \begin{overpic}[width=0.996\textwidth]{img/sup_1.pdf}
  \put(25,2){\small{Input}}
  \put(94,2){\small{VFN~\cite{chen2017learning}}}
  \put(162,2){\small{GAIC~\cite{zeng2019reliable}}}
  \put(236,2){\small{CGS~\cite{li2020composing}}}
  \put(289,2){\small{Zhong~\etal~\cite{zhong2021aesthetic}}}
  \put(368,2){\small{Jia~\textit{et~al.}~\cite{jia2022rethinking}}}
  \put(452,2){\small{Ours}}
\end{overpic}
\vspace{1mm}
\caption{Qualitative comparison with other methods on FLMS ~\cite{fang2014automatic} dataset.}
\label{fig:NAIC_compare_flms}
\end{figure*}

\begin{figure*}[htp]
\centering
 \begin{overpic}[width=0.996\textwidth]{img/sup.pdf}
  \put(25,2){\small{Input}}
  \put(95,2){\small{VFN~\cite{chen2017learning}}}
  \put(162,2){\small{GAIC~\cite{zeng2019reliable}}}
  \put(235,2){\small{CGS~\cite{li2020composing}}}
  \put(289,2){\small{Zhong~\etal~\cite{zhong2021aesthetic}}}
  \put(369,2){\small{Jia~\textit{et~al.}~\cite{jia2022rethinking}}}
  \put(452,2){\small{Ours}}
\end{overpic}
\vspace{1mm}
\caption{Qualitative comparison with other methods on GAICD ~\cite{zeng2019reliable} dataset.}
\label{fig:NAIC_compare}
\end{figure*}

{\small
\bibliographystyle{ieee_fullname}
\bibliography{egbib}
}